\magnification1200


\vskip 2cm
\centerline
{\bf  An $E_{11}$  invariant gauge fixing}
\vskip 1cm
\centerline{ Michaella Pettit  and Peter West}
\centerline{Department of Mathematics}
\centerline{King's College, London WC2R 2LS, UK}
\vskip 2cm
\leftline{\sl Abstract}
We consider the non-linear realisation of the semi-direct product of $E_{11}$ and its vector representation which leads to a  spacetime with tangent group that is the Cartan involution invariant subalgera of $E_{11}$.  We give an alternative derivation of the invariant tangent space metric that this space-time possesses and compute this metric  at low levels in eleven, five  and four dimensions. We show that one can gauge fix the non-linear realisation in an $E_{11}$   invariant manner.

\vskip2cm
\noindent

\vskip .5cm

\vfill
\eject
{\bf 1 Introduction}
\medskip
One of the most surprising discoveries in the construction of supergravity theories was that the $N=8$ supergravity theory in four dimensions possesses  an $E_7$ symmetry [1]. It was then found that the maximal supergravities in dimensions $D\le 9$ possesses an $E_{11-D}$ symmetry [2] and that the IIB supergravity has an SL(2,R) symmetry [3]. It was universally thought that these symmetries, with the exception of the SL(2,R) symmetry  of IIB, were a quirk of the dimensional reduction procedure that can be used to relate the different theories. However, it was proposed in 2001 [4] that these symmetries lifted  and that there existed a theory in eleven dimensions that has an $E_{11}$ symmetry that contained the maximal supergravity theory. It was proposed in 2003 [5] that this symmetry also acted on the spacetime and that one should consider the non-linear realisation of the semi-direct product of $E_{11}$ with its vector representation, denoted $E_{11}\otimes_s l_1$. This theory possesses an infinite number of fields which live on an infinite dimensional spacetime.
\par
The non-linear realisation automatically provides some of the geometrical structures for the spacetime that it  contains. The tangent space group is the Cartan involution invariant subgoup of $E_{11}$, denoted $I_c(E_{11})$,  and it possesses a vielbein constructed from the fields of the theory. At the lowest level these are respectively  the Lorentz group and the usual vielbein  of general relativity.  In this paper we will explain   that the tangent space also possesses a metric invariant under $I_c(E_{11})$ analogous to the Minkowski metric that exists in the tangent space of general relativity. We evaluate this metric at low levels in eleven, five and four dimensions. 
\par
The existence of such an invariant tangent space metric follows from the known literature on Kac-Moody algebras. Indeed it is known that any highest weight representation of any Kac-Moody algebra possesses an invariant metric [20]. As the tangent vectors of the space-time belong to the vector representation it then follows  from this result that they possess an $I_c(E_{11})$ invariant metric. The argument given in this paper has the advantage that it leads to detailed formulae that are useful for the construction of the non-linear realisation.
As an application we construct an
$E_{11}\otimes_s l_1$ invariant set of equations that can be interpreted as an $E_{11}$ covariant gauge fixing conditions.
\par
Although we will mainly find the results for $E_{11}\otimes_s l_1$ they can easily be generalised  to  the non-linear realisation of any algebra of the form $G^{+++}\otimes_s l_1$ and we will also apply the method to the non-linear realisation
$A_1^{+++}\otimes_s l_1$ which is associated with gravity.

 The commutators of the  algebra $E_{11}\otimes_s l_1$ can be written in the form
$$
[R^{\underline \alpha} , R^{\underline \beta} ]= f^{\underline \alpha \underline \beta}{}_{\underline \gamma} R^{\underline \gamma}, \quad
[R^{\underline \alpha} , l_A]= -(D^{\underline \alpha} )_A{}^B l_B
\eqno(1.1)$$
where $R^{\underline \alpha}$ are the generators of $E_{11}$ and $l_A$ are the generators belonging to the $l_1$ representation. We assume that the $l_1$ generators commute. The Jacobi identities imply that the matrices $(D^{\underline \alpha} )_A{}^B$ are a representation of the $E_{11}$ algebra,
$$[D^{\underline \alpha}, D^{\underline \beta }]= f^{\underline \alpha \underline \beta}{}_{\underline \gamma} D^{\underline \gamma}
\eqno(1.2)$$
\par
We first very briefly recall how the non-linear realisation of $E_{11}\otimes_sl_1$ is constructed. A more complete  account can be found in the review of reference [6] where further references can be found.
We first consider a group element  $g\in E_{11}\otimes_sl_1$ that  can be written as
$$
g=g_lg_E
\eqno(1.3)$$
In this equation  $g_E$ is a group element of $E_{11}$ and  it can be written in the form
$g_E=e^{A_{\underline \alpha} R^{\underline \alpha}}$ where   $A_{\underline\alpha}$ will become  the fields in the non-realisation. The group element   $g_l$ is formed from the generators of the $l_1$ representation and so has the form $e^{z^A L_A} $ where $z^A$ are the coordinates of the generalised space-time. The fields $A_{\underline\alpha}$ depend on the coordinates $z^A$. 
\par
The non-linear realisation is, by definition, invariant under the transformations
$$
g\to g_0 g, \ \ \ g_0\in E_{11}\otimes _s l_1,\ \ {\rm as \  well \  as} \ \ \ g\to gh, \ \ \ h\in
I_c(E_{11})
\eqno(1.4)$$
The group element $g_0\in E_{11}$ is a rigid transformation, that is, it is  a constant,  while the group element $h$ belongs to the Cartan involution subalgebra of $E_{11}$, denoted $I_c(E_{11})$. This latter transformation  is a local
transformation, that is,  it depends on the generalised space-time with coordinates $z^A$. As the generators in $g_l$ form a representation of $E_{11}$ the above transformations for $g_0\in E_{11}$ can be written as
$$
g_l\to g_0 g_lg_0^{-1}, \quad g_E\to g_0 g_E\quad {\rm and } \quad g_E\to g_E h
\eqno(1.5)$$
\par
The Cartan involution  will play an important part in this paper.
It is an automorphism of the algebra, that is,  $I_c(AB)= I_c (A)I_c(B)$ for any two elements $A$ and $B$ of the Lie algebra and an involution, meaning that   $I_c^2(A)= A$. It takes positive root generators to negative root generators and its action can be taken to be
$$
I_c(R^{\underline \alpha}) = - R^{-\underline \alpha}
\eqno(1.6)$$
for any root $\alpha$. The Cartan Involution subalgebra is generated by
$R^{\underline \alpha} - R^{-\underline \alpha} $. For the finite dimensional semi-simple Lie groups the Cartan Involution invariant subalgebra is  the maximal compact subalgebra; for example for SL(2,R) the Cartan Involution subalgebra is SO(2), while for $E_8$ it is SO(16). Given the positive root generators one can define the negative root generators using equation (1.6) and while it is simpler to take a minus sign on the right-hand side for all generators the subject has developed in such a way that one finds a plus sign for even level  generators. For simplicity when explaining the general theory in this and the next section we take equation (1.6) to hold.
\par
The Cartan involution  of equation (1.6) leads to a theory that lives in a  spacetime  with Euclidean signature, but by altering the signs in the Cartan involution one can construct a theory with a  Minkowski signature when the generalised spacetime is truncated to our usual spacetime [21]. To avoid having to take account of these signs as they propagate through the construction it is often easier to work in  Euclidean signature and then change to Minkowski signature at the end.
\par
The dynamics of the non-linear realisation $E_{11}\otimes_s l_1$  is just a  set of equations of motion, that are invariant under the transformations of equation (1.4). The reader is free to achieve this goal in any way but the usual method is  to construct the dynamics of the
the non-linear realisation using the  Cartan forms  which are given by
$$
{\cal V}\equiv g^{-1} d g= {\cal V}_E+{\cal V}_l,
\eqno(1.7)$$
where
$$
{\cal V}_E=g_E^{-1}dg_E\equiv dz^\Pi G_{\Pi, \underline \alpha} R^{\underline \alpha},
 \eqno(1.8)$$
belongs to the $E_{11}$ algebra and are the Cartan form for $E_{11}$
and other part of the Cartan form which contains the generators of the $l_1$ representation is given by
 $$
{\cal V}_l= g_E^{-1}(g_l^{-1}dg_l) g_E= g_E^{-1} dz\cdot l g_E\equiv
dz^\Pi E_\Pi{}^A l_A 
\eqno(1.9)$$
\par
While  both ${\cal V}_E$ and ${\cal V}_l$ are invariant under rigid transformations,  under the local transformations of equation (1.4) they change as
$$
{\cal V}_E\to h^{-1}{\cal V}_E h + h^{-1} d h\quad {\rm and }\quad
{\cal V}_l\to h^{-1}{\cal V}_l h
\eqno(1.10)$$
\par
The action of a finite transformation of the $E_{11}$ on the generators of the $l_1$ representation is, by definition,  given by
$$
U(k)( l_A)\equiv k^{-1} l_A k= D(k)_A{}^B l_B, \quad k\in E_{11}
\eqno(1.11)$$
where  $D(k)_A{}^B $  is the
matrix representative of the finite transformation.  It follows from equation (1.1) that $D(k)_A{}^B =(e^{a_{\underline \alpha}D^{\underline \alpha}})_A{}^B$ when $k= e^{a_{\underline \alpha}R^{\underline \alpha}}$.
Examining equation (1.9), and recalling equation (1.1),  we recognise
${ E}_\Pi{}^{A} $ as the representation  matrix $D(g_E)_\Pi{}^A$, and so
${ E}_\Pi{}^A = D(g_E)_\Pi{}^A= (e^{A_{\underline \alpha}D^{\underline \alpha}})_\Pi{}^A$. The indices on this last object are   labelled according to the role which they will play later in the  theory that emerges from the non-linear realisation. 
\par
It follows from equation (1.5) that the coordinates are inert under the local transformations but transform under the rigid  transformations as
$$
z^A l_A\to z^{A\prime} l_A=g_0 z^Al_A g_0^{-1} = z^\Pi D(g_0^{-1})_\Pi {}^Al_A
\eqno(1.12)$$
where $D(g_0^{-1})_\Pi {}^A$ is the representation matrix of equation (1.11).
When  written  in matrix form the differential  transformations act as 
$dz^T\to dz^{T\prime}= dz ^T D(g_0^{-1})$.   On the other hand the derivative
$\partial_\Pi\equiv {\partial\over \partial z^\Pi}$ in the generalised space-time
transforms as $\partial_\Pi^\prime= D(g_0)_\Pi{}^\Lambda \partial_\Lambda$. Again the use of different indices will correspond to the interpretation of these objects that follows from the non-linear realisation.
\par
The Cartan form is inert under the rigid $g_0$ transformations of equation (1.4). However,  the generalised vielbein  $E_\Pi{}^A$, defined in equation (1.9), occurs in the combination $dz^\Pi E_\Pi{}^A$ and as such it undergoes a transformation induced by the rigid $g_0$ transformation of equation (1.12) on $dz^\Pi$, which also  acts on the  $\Pi$ index of $E_\Pi{}^A$.
Under    a local $I_c(E_{11})$ transformation, $E_\Pi{}^A$ transforms
 on its $A$ index  as governed by  equation (1.10).    We may summarise these two results as
$$
{ E}_\Pi{}^{A\prime} =
D(g_0)_\Pi{}^\Lambda { E}_\Lambda{}^{B}D(h)_B{}^A \quad {\rm or}\quad \quad  (E^{-1})_A{}^{\Pi\prime}= D(h^{-1})_A{}^B (E^{-1})_B{}^\Lambda
D(g_0^{-1})_\Lambda{}^\Pi
\eqno(1.13)$$
\par
Thus   $E_\Pi{}^A$ transforms
under a local $I_c(E_{11})$ transformation on its $A$ index and by a rigid $E_{11}$ induced
coordinate transformation of the  space-time on its $\Pi$ index.  Thus  we can interpretation of ${ E}_\Pi{}^{A}$ as a  vielbein of the space-time which possesses a
tangent space with  the tangent group $I_c(E_{11})$. 

\medskip
{\bf 2 The tangent space metric}
\medskip
As we have discussed  above, the  spacetime, contained in the $E_{11}\otimes_s l_1$ non-linear realisation,  possesses a tangent space with a tangent group $I_c(E_{11})$ and so we can consider tangent space objects, such as  $V^A$,  whose transformation is defined to be  [7]
$$
V^{A\prime}  l_A= h^{-1}V^A l_A h  \quad \quad h\in I_c(E_{11})
\eqno(2.1)$$
Such infinitesimal transformations for the group element
$h= 1+a_{\underline \alpha}(R^{\underline \alpha}- R^{-\underline \alpha})$ can be written as
$$
V^{A\prime}= V^B D(h)_B{}^A= V^A+ V^B (D^{\underline \alpha}- D^{-\underline \alpha}) _B{}^A a_{\underline \alpha}+ \dots
\eqno(2.2)$$
\par
One can use the vielbein of equation (1.9)  to convert world to tangent vectors and tensors and vice-versa, that is, $V^\Pi =
V^A (E^{-1})_A {}^\Pi$. The world object $V^\Pi$ is inert under local  $I_c(E_{11})$ transformations but it does transform under local transformations as given in equation (1.13) in reference [7].
\par
We will now show that the generalised tangent space possesses a metric which is invariant under the tangent space group $I_c(E_{11})$. This is the analogue of the tangent space metric of general relativity, that is, the Minkowski metric, which is of course invariant under the  local Lorentz transformations of general relativity found in the vielbein formalism. The Cartan involution will play an important role in this construction.
\par
 The Cartan involution swaps negative with positive roots, and so  the corresponding generators. As such it exchanges raising with lowering generators when they act on a representation and so it will take a highest weight representation into a lowest weight representation and visa-versa.
Despite being sometimes  carelessly labelled,  the $l_1$ representation is in fact a lowest weight representation and under the action of $I_c$ it will lead to a highest weight representation. We denote this latter representation by $\bar l_1$ and  its   components by $\bar l^A$.   The action of $I_c$ on the $l_1$ representation can be written as
$$
I_c (l_A)= -J_{AB}^{-1} \bar l^B
\eqno(2.3)$$
where $J_{AB}$ is a constant matrix. One can take $J_{AB}= \delta_{AB}$ but in certain cases it is desirable to have it be a non-trivial constant invertible matrix.
\par
We can write the commutators of the generators of $E_{11}$ with the generators of $\bar l_1$ as
$$
[R^{\underline \alpha } ,\bar l^A]= \bar l^B (\bar D^{\underline \alpha }) _B{}^A
\eqno(2.4)$$
where the matrices $(\bar D^{\underline \alpha }) $ obey the $E_{11}$ commutation relation
$
[(\bar D^{\underline \alpha }) , (\bar D^{\underline \beta }) ]= f^{\underline \alpha \underline \beta}{}_{\underline \gamma} (\bar D^{\underline \gamma }) $ as a consequence of the Jacobi identities.
Acting with the Cartan involution on equation (2.3) and using equations (1.1) and (2.4)  we conclude that
$$
\bar D^{\underline \alpha } = (J D^{-\underline \alpha } J^{-1})^T
\eqno(2.5)$$  
\par
From the $l_1$ and $\bar l_1$ representations we can construct a map $l_1\otimes \bar l_1 \to C$ which  is $E_{11}$ invariant. Let us  denote the  value of the map   by $N_A{}^B\equiv (l_A, \bar l^B )$; whereupon being  invariant means that
$([R^{\underline \alpha },l_A ], \bar l^B )+ (l_A, [ R^{\underline \alpha }, \bar l^B ])=0$,  or equivalently, that 
$$
D^{\underline \alpha }N=N\bar D^{\underline \alpha }
\eqno(2.6)$$
Using equation (2.5) to eliminate $\bar D^{\underline \alpha }$ we find that
$$
D^{\underline \alpha } K= K (D^{- \underline \alpha } )^T
\eqno(2.7)$$
where the object $K$ is defined by
$$K=N(J^{-1})^T
\eqno(2.8)$$
The object $K$ will  turn out to be the invariant metric we seek.
\par
Taking the double transpose of equation (2.7) and  we find that
$$
D^{\underline \alpha }= ((D^{\underline \alpha } )^T)^T= (K^{-1}(D^{-\underline \alpha })K)^T=
(K)^T  (D^{-\underline \alpha })^T(K^{-1})^T=
(K)^T K^{-1} (D^{\underline \alpha }) K(K^{-1})^T
\eqno(2.9)$$
Rewriting this equation we have that
$$
K(K^{-1})^T D^{\underline \alpha }= D^{\underline \alpha }K(K^{-1})^T
\eqno(2.10)$$
Schur's lemma then allows us to  conclude that $ K( K^{-1})^T=1 $ and so
$$
K= K^T
\eqno(2.11)$$
\par
Using this last result we can rewrite equation (2.7) in the form
$$
D^{\underline \alpha}K= (D^{-\underline \alpha }K)^T
\eqno(2.12)$$
which  tells us how the representation matrix $D^{\underline \alpha}K$ behaves under transposition.  As a result 
$$
(D^{\underline \alpha}\pm D^{-\underline \alpha})K= \pm ((D^{\underline \alpha}\pm D^{-\underline \alpha})K)^T
\eqno(2.13)$$
We note that $(D^{\underline \alpha}- D^{-\underline \alpha})K$ is an antisymmetric matrix. An obvious consequence is that
$$
(D^{\underline \alpha}- D^{-\underline \alpha})K
+ K (D^{\underline \alpha}-  D^{-\underline \alpha})^T=0
\eqno(2.14)$$
As  $(D^{\underline \alpha}- D^{-\underline \alpha})$ is an $I_c(E_{11})$ transformation in the vector representation we conclude that  $K$ is an invariant tensor of $I_c(E_{11})$ and so it is the invariant metric that we seek. 
\par
The tangent space object  $V^A$,  transforms as in equation (2.2) and as a result  we find,  using equation (2.13),  that the quantity
$$
\Delta = V^A K_{AB} V^B =V^T K V
\eqno(2.15)$$
is invariant.
\par
We can use the metric $K_{AB}$ to lower the tangent index on the tangent vector $V^A$ as follows  $W_A= V^B K_{BA}$ and one finds, using equation (2.25), that
$$
\delta W_A = - (D^{\underline \alpha}- D^{-\underline \alpha})_A{}^B  a_{\underline \alpha} W_B
\eqno(2.16)$$
\par
We note that if we take the tangent objects to be infinitesimal distances, that is, $V^A = dz^{\Pi} E_{\Pi}{}^A$ we can write an invariant distance in the generalised spacetime which is given by
$$
{\rm d}s^2= dz^{\Pi} g_{\Pi\Lambda}dz^{\Lambda} , \quad {\rm or}\quad
g_{\Pi\Lambda}=  E_{\Pi}{}^A K_{AB}  E_{\Lambda}{}^B
$$
\par
Although we have explained the construction of the tangent space metric  for the non-linear realisation $E_{11}\otimes_s l_1$ the method used  applies equally well to the non-linear realisation of any algebra of the form $G^{+++}\otimes_s l_1$ and indeed any algebra which is the semi-direct product of any Lie   algebra $G$ with one of its representations $l$, that is,  $G\otimes _s l$. 
\medskip


\medskip
{\bf 3. The invariant tangent space metric in various dimensions}
\medskip
We now apply the theory of section two to find the invariant tangent space metric in the $E_{11}\otimes_s l_1$ non-linear realisation in eleven, five  and four   dimensions and also the non-linear realisation of the algebra $A_1^{+++}\otimes l_1$ which leads to gravity. However, we will begin by considering the non-linear realisation of the simpler algebra $IGL(D)\equiv GL(D)\otimes_s T^D$.   
\medskip
{\bf 3.1 IGL(d) and Gravity}
\medskip
The discussion of section two is a bit abstract so we first apply it in the most well known context of gravity which emerges as the   non-linear realisation of the group $IGL(D)$ [8,9] provided one chooses appropriately  the constants that  appear in the resulting equations of motion. In this case the $l_1$ representation is the vector representation of GL(D), denoted $T^D$ and whose  corresponding generators are just the usual space-time translations  denoted by $P_a$. We denote the generators of GL(D) by $K^a{}_b$ and the semi-direct product algebra $IGL(D)
\equiv GL(d)\otimes_s T^D$ is given by
$$
[K^a{}_b,K^c{}_d]= \delta_b^c K^a{}_d- \delta_a^d K^c{}_b, \quad [K^a{}_b, P_c]=-\delta^a_c P_b
\eqno(3.1.1)$$
Given the vectors $e_a, \ a=1,2,\ldots , D$,  which obey $(e_a , e_b )=\delta_{ab}$, the roots of $GL(D)$ can be written as  $e_a-e_b$  and so they are labelled by a pair of integers $(a,b)$. The simple roots are given by $\alpha_a= e_a-e_{a+1}$. The action of the Cartan involution takes the root corresponding to $(a,b)$ to the root corresponding to $(b,a)$ and the effect on the generators is
$I_c(K^a{}_b)= -K^b{}_a$. The Cartan involution invariant subalgebra is generated by
$ \delta _{ae}K^e{}_b-\delta _{be}K^e{}_a$ giving  the  algebra SO(D). As explained above, if we scatter some signs in the action of the Cartan involution we would find the Minkowski metric  $\eta_{ae}$ rather than the Euclidean metric $\delta_{ab}$ in the last equation.
\par
The $l_1$ representation of equation (3.1.1) has the matrix  $(D^a{}_b)_c{}^d= \delta^a_c\delta_b^d$.  From our discussion just above we find that if  $D^{\underline \alpha }$ corresponds to $D^a{}_b$ then $D^{- \underline \alpha }$ corresponds to $D^b{}_a$.
The Cartan involution takes us from the $l_1$ representation to a representation $\bar l_1$ which contains  the generators
$\bar P^a$. Its   action is given by $I_c(P_a)= -\bar P^a$,  which means we have taken   $J_{ab}=\delta_{ab}$ in equation (2.3),  and the commutator of equation (2.4) takes the form
$$
[K^a{}_b, \bar P^c]=\delta_b^c \bar P^a
\eqno(3.1.2)$$
from which we identify  $(\bar D^a{}_b)_d{}^c= \delta ^a_d\delta^c_b$.
\par
Denoting the scalar product between the $P_a$ and $\bar P^a$ representations 
by   $N_a{}^b = (P_a,\bar P^b)$ we find that it is SL(D) invariant if $([K^a{}_b , P_c], \bar P_d)+ (P_c, [K^a{}_b , \bar P^d)=0$.
Using equations (3.1.1) and (3.1.2) we find that $N_a{}^b = \delta_a^b$.  
As a result  we find that the invariant metric is $K_{ab}= \delta_{ab}$.
It obviously satisfies equation (2.11), that is,  $K^T=K$. The identity of equation (2.12) is satisfied and takes the form $D^a{}_b = (D^b{}_a)^T$.Equation (2.14) is just the well known fact that the tangent space metric $\delta_{ab}$ is preserved by SO(D) transformations. Of course we have not found any result which is new in this last section, however, we have derived  them in a way which applies to the  more sophisticated  spacetimes which arise in the non-linear realisations being considered in this paper.

\medskip
{\bf 3.2 $E_{11}$ in eleven dimensions}
\medskip
 The $E_{11} \otimes_s l_1$ theory has been discussed extensively, in particular, the book of reference [10] gives a detailed account of the algebra and its mathematical underpinning while the papers  [4,11, 12,13] contain a  discussion of the physical application. We obtain the eleven dimensional theory by deleting node 11 from the Dynkin diagram of $E_{11}$:
$$ \matrix{ & & & & & & & & & & & & & & & \otimes & 11 & & & \cr
& & & & & & & & & & & & & & & | & & & & \cr
& \bullet & - & \bullet & - & \bullet & - & \bullet & - & \bullet & - & \bullet & - & \bullet & - & \bullet & - & \bullet & - & \bullet \cr
& 1 & & 2 & & 3 & & 4 & & 5 & & 6 & & 7 & & 8 & & 9 & & 10 \cr} $$
and we decompose the $E_{11} \otimes_s l_1$ algebra into representations of the remaining algebra which is $GL(11)$. Up to level 3, the non-negative level generators of $E_{11}$ are:
 
$$ K^a{}_b; \ R^{a_1a_2a_3}; \ R^{a_1\ldots a_6}; \ R^{a_1\ldots a_8,b} . \eqno{(3.2.1)} $$
The negative level generators of $E_{11}$, up to level 3, are:
 
$$ R_{a_1a_2a_3} ; \ R_{a_1\ldots a_6} ; \ R_{a_1\ldots a_8,b} . \eqno{(3.2.2)} $$
The action of the Cartan involution subalgebra is defined in equation (2.6). In eleven dimensions, it is given by:
 
$$ I_c(K^a{}_b) = -K^b{}_a, \ \ \ I_c(R^{a_1a_2a_3}) = -R_{a_1a_2a_3}, $$
$$ I_c(R^{a_1\ldots a_6}) = R_{a_1\ldots a_6}, \ \ \ I_c(R^{a_1\ldots a_8,b}) = -R_{a_1\ldots a_8,b} .
\eqno{(3.2.3)} $$
The reader will notice that one of the above relations has a plus sign rather than the universal minus sign of equation (1.6). The sign one chooses  is a matter of convention as it is used to define the way the negative root generators occur in the algebra. In order to agree with the many the previous $E_{11}$  papers  we adopt this plus sign and modify the general theory of section two to take account of it.
Up to level 3, the $l_1$ representation contains the following generators:
 
$$P_a; \ Z^{a_1a_2}; \ Z^{a_1\ldots a_5}; \ Z^{a_1\ldots a_8}, \ Z^{a_1\ldots a_7,b}, \ldots , \eqno{(3.2.4)} $$
and the $\bar{l}_1$ generators, down to level -3, are:
 
$$ \bar{P}^a; \ \bar{Z}_{a_1a_2}; \ \bar{Z}_{a_1\ldots a_5}; \ \bar{Z}_{a_1\ldots a_8}, \ \bar{Z}_{a_1\ldots a_7,b}, \ldots . \eqno{(3.2.5)} $$
The  commutators of the $E_{11} \otimes_s l_1$ algebra are given the book of reference [10]. The commutators of the $l_1$ and $\bar{l}_1$ generators  with the $E_{11}$ generators  are given in Appendix A.1.
\par
The action of the Cartan involution on the $l_1$ generators  is given by
$$
I_c(P_a) = -\bar{P}^a, \ I_c(Z^{a_1a_2}) = -\bar{Z}_{a_1a_2}, \ I_c(Z^{a_1\ldots a_5}) = -\bar{Z}_{a_1\ldots a_5} , $$
$$ I_c(Z^{a_1\ldots a_8}) = - \bar{Z}_{a_1\ldots a_8}, \ I_c(Z_{a_1\ldots a_7,b}) = -\bar{Z}_{a_1\ldots a_7,b} .
\eqno{(3.2.6)} $$
From equation (2.15), we note that this action of the Cartan involution subalgebra defines the object $J_{AB}$ to be given by $J_{AB} = \delta_{AB}$.
\par
 The group element $g=g_lg_A$ in the eleven dimensional theory can be written as
$$
g_l = \exp{(x^aP_a + x_{a_1a_2}Z^{a_1a_2} + x_{a_1\ldots a_5}Z^{a_1\ldots a_5} + x_{a_1\ldots a_8}Z^{a_1\ldots a_8} + x_{a_1\ldots a_7,b}Z^{a_1\ldots a_7,b})} \ldots
\eqno(3.2.7)$$
$$g_A = \exp(h_a{}^bK^a{}_b)\exp(A_{a_1\ldots a_8,b}R^{a_1\ldots a_8,b})\exp(A_{a_1\ldots a_6}R^{a_1\ldots a_6})\exp(A_{a_1a_2a_3}R^{a_1a_2a_3})
\ldots
\eqno(3.2.8)$$
where  the parameters in $g_l$
$$
x^a ; \ x_{a_1a_2} ; \ x_{a_1\ldots a_5} ;\ x_{a_1\ldots a_8} , \ x_{a_1\ldots a_7,b} \ldots ,
\eqno(3.2.9)$$
can be interpreted as the coordinates of the  space-time.
While the parameters of $g_E$ are the fields
$$
h_a{}^b ;\ A_{a_1a_2a_3} ; \ A_{a_1\ldots a_6} ; \ A_{a_1\ldots a_8,b} , \ldots .
\eqno(3.2.10)$$
which  live  on the  space-time.
These above explicitly listed fields correspond to the graviton, the three and six form gauge fields, and the dual graviton, respectively.
In writing the group element $g_E$ of equation (3.2.10) we have used  the
local symmetry $I_c(E_{11})$ to set to zero all the negative level fields.
\par
The tangent object  introduced in section 2 has the components
$$ V^A = (T^a, T_{a_1a_2}, T_{a_1...a_5}, T_{a_1...a_8}, T_{a_1...a_7,b}, \ldots ) , \eqno{(3.2.11)} $$
in eleven dimensions.
The level zero transformations of $I_c(E_{11})$ are just local Lorentz transformations. The  infinitesimal transformation $h\in I_c(E_{11})$ which contains generators at  levels plus and minus one is given in eleven dimensions by
$$
h = 1 - \Lambda_{a_1a_2a_3}(R^{a_1a_2a_3} - \eta ^{a_1 b_1}\eta ^{a_2 b_2}
\eta ^{a_3 b_3}
R_{b_1b_2b_3}) .
\eqno(3.2.12) $$
By taking multiple commutators of the generator that appears in this last equation  one finds all the generators of  $I_c(E_{11})$. It follows that if a theory is invariant under these $I_c(E_{11})$ transformations it is invariant under all $I_c(E_{11})$ transformations.
\par
The  $I_c(E_{11})$ transformations of the tangent objects are given in equation (2.2) and using the commutators of $E_{11}$ generators and the $l_1$ generators given in the appendix A.1 we find that, up to the level considered
 $$ \delta T^a = -6\Lambda^{a_1a_2a}T_{a_1a_2} , $$
$$ \delta T_{a_1a_2} = 3\Lambda_{a_1a_2 b}T^b - 60\Lambda^{b_1b_2b_3}T_{b_1b_2b_3a_1a_2} , $$
$$ \delta T_{a_1\ldots a_5} = \Lambda_{[a_1a_2a_3}T_{a_4a_5]} + 42\Lambda^{b_1b_2b_3}T_{b_1b_2b_3a_1\ldots a_5}
-378 \Lambda^{b_1b_2b_3}T_{b_1b_2b_3[a_1\ldots a_4,a_5]},  $$
$$ \delta T_{a_1\ldots a_8} = -\Lambda_{[a_1a_2a_3}T_{a_4\ldots a_8]} ,$$
$$ \delta T_{a_1\ldots a_7,b} = T_{[a_1\ldots a_5}\Lambda_{a_6a_7]b} - T_{[a_1\ldots a_5}\Lambda_{a_6a_7b]} ,
\eqno(3.2.13)$$
where indices are raised and lowered with the Minkowski metric.
\par
We now want to find the metric which is invariant under the action of the Cartan involution subalgebra for the eleven dimensional theory. This metric was defined in general in equation (2.8). We will now derive it in two ways. Firstly, we will  derive it using the general theory of section  two, that is, we will construct the invariant map $N_A{}^B$, discussed above equation (2.6), and then find the invariant tangent space metric using equation (2.8).
We will also find the invariant metric by using the above transformations of the tangent objects under $I_c(E_{11})$ transformations given in equation (3.2.13) and then find the invariant metric, by explicit construction.
\par
We begin by using the first method, that is, by first finding the invariant $N_A{}^B \equiv (l_A,\bar{l}^B)$  constructed from the $l_1$ and $\bar l_1$ representations using the fact that it is invariant under $E_{11}$ transformations. As discussed in section 3.1 we find that $N_a{}^b = \delta^b_a = (P_a,\bar{P}^b)$ using the fact that $N_a{}^b$  is invariant under SL(11) transformations.
To find $(Z^{a_1a_2},\bar{Z}_{b_1b_2})$, we note that  $E_{11}$ invariance of the scalar product implies  that
$$ ([R^{a_1a_2a_3},P_b],\bar{Z}_{c_1c_2}) + (P_b, [R^{a_1a_2a_3},\bar{Z}_{c_1c_2}]) = 0 .
\eqno(3.2.14)$$
and so
$$ 3\delta_b^{[a_1}(Z^{a_2a_3]},\bar{Z}_{c_1c_2}) - 6\delta_{c_1c_2}^{[a_1a_2}(P_b,\bar{P}^{a_3]}) = 0 ,
\eqno(3.2.15)$$
We conclude that [14]
$$ (Z^{a_1a_2},\bar{Z}_{b_1b_2}) = 2\delta^{a_1a_2}_{b_1b_2} .
\eqno(3.2.16)$$
\par
Proceeding in the same way we find that
$$ N_A{}^B = \pmatrix{ \delta_a^b & 0 & 0 & 0 & 0 \cr
0 & 2\delta^{a_1a_2}_{b_1b_2} & 0 & 0 & 0 \cr
0 & 0 & 5! \delta^{a_1\ldots a_5}_{b_1\ldots b_5} & 0 & 0 \cr
0 & 0 & 0 & 7! \delta^{a_1\ldots a_8}_{b_1\ldots b_8} & 0 \cr
0 & 0 & 0 & 0 & 9(7!)\delta^{a_1\ldots a_7,a}_{b_1\ldots b_7,b} \cr} \eqno{(3.2.17)}$$
where $\delta^{a_1\ldots a_7,a}_{b_1\ldots b_7,b} = \delta^{a_1\ldots a_7}_{b_1\ldots b_7}\delta^a_b - \delta ^{a_1\ldots a_7a}_{b_1\ldots b_7b}$ and $\delta^{a_1\ldots a_n}_{b_1\ldots b_n} = \delta^{[a_1}_{b_1}\ldots\delta^{a_n]}_{b_n}$.
\par
The $I_c(E_{11})$ invariant metric was given  in equation (2.8) by $K=N(J^{-1})^T$, where we have omitted the indices.
As we noted above in the eleven dimensional theory, we have $J_{AB}=\delta_{AB}$. As a result we find, up to the levels considered,  that
$$K_{AB} = \pmatrix{ \eta_{ab} & 0 & 0 & 0 & 0 \cr
0 & 2\delta^{a_1a_2,b_1b_2} & 0 & 0 & 0 \cr
0 & 0 & 5! \delta^{a_1\ldots a_5,b_1\ldots b_5} & 0 & 0 \cr
0 & 0 & 0 & 7!\delta^{a_1\ldots a_8,b_1\ldots b_8} & 0 \cr
0 & 0 & 0 & 0 & 9(7!) \delta^{a_1\ldots a_7,a,b_1\ldots b_7,b} \cr} \eqno{(3.2.18)} $$
We have defined $\delta^{a_1\ldots a_7,a,b_1\ldots b_7,b} = \delta^{a_1\ldots a_7,b_1\ldots b_8}\delta^{ab} - \delta^{a_1\ldots a_7a,b_1\ldots b_7b}$ and $\delta^{a_1\ldots a_n}_{b_1\ldots b_n} = \delta^{[a_1}_{b_1}\delta^{a_2}_{b_2}\ldots \delta^{a_n]}_{b_n}$ as usual.
\par
We now recover the same result using the second method, that is,  by using  the transformations of the tangent space vectors found in equation (3.2.13).
We simply write down the most general expression that is quadratic in the tangent objects and determine the coefficients by explicitly testing if it is invariant  under  these transformations.   The result is
$$ \Delta =
T^aT_a
+ 2T^{a_1a_2}T_{a_1a_2}
+ 5! T^{a_1\ldots a_5}T_{a_1\ldots a_5}
+ 7! T^{a_1\ldots a_8}T_{a_1\ldots a_8}
$$
$$
+ 9(7!) T^{a_1\ldots a_7,b}T_{a_1\ldots a_7,b} \ldots
\eqno{(3.2.19)}$$
which agrees with the above found expression for $K_{AB}$ of equation (3.2.18).


\medskip
{\bf
3.3 $E_{11}$ in 5 dimensions}
\medskip
The five dimensional theory has been discussed in [13,15]. We obtain the five dimensional theory by deleting node 5 from the Dynkin diagram of $E_{11}$:
$$ \matrix{ & & & & & & & & & & & & & & & \bullet & 11 & & & \cr
& & & & & & & & & & & & & & & | & & & & \cr
& \bullet & - & \bullet & - & \bullet & - & \bullet & - & \otimes & - & \bullet & - & \bullet & - & \bullet & - & \bullet & - & \bullet \cr
& 1 & & 2 & & 3 & & 4 & & 5 & & 6 & & 7 & & 8 & & 9 & & 10 \cr} $$
 and  decomposing  the $E_{11} \otimes_s l_1$ algebra into representations of the remaining algebra which is $GL(5)\times E_6$. In this paper we are interested in the tangent space which has the tangent group $I_c(E_{11}) $  and as a result we will decompose the algebra $E_{11}\otimes _s l_1$  into representations of the level zero tangent space group, that is,
$I_c(GL(5) \otimes E_6) = SO(5) \otimes Usp(8)$ rather than $GL(5) \otimes E_6$. The five dimensional theory in this decomposition  was discussed in [13]. 
\par
 Up to level 3, the non-negative level generators of  $E_{11}$ algebra when decomposed in this way are
 $$ K^a{}_b, \ R^{\alpha_1\alpha_2}, R^{\alpha_1\ldots \alpha_4}; \ R^{a\alpha_1\alpha_2} ; \ R^{a_1a_2}{}_{\alpha_1\alpha_2} ; \ R^{a_1a_2a_3\alpha_1\alpha_2} , \ R^{a_1a_2a_3\alpha_1\ldots \alpha_4}, R^{a_1a_2,b}. \
\eqno{(3.3.1)} $$
where $\alpha_1 ,  \alpha_2 , \ldots = 1,\ldots, 8$, and we use the Usp(8) invariant, antisymmetric  metric $\Omega_{\alpha_1\alpha_2}$ and its inverse to raise and lower indices as follows $T^\beta = \Omega ^{\beta \gamma }T_\gamma$,  $T_\alpha= \Omega _{\alpha\beta}T^\beta$  and so
$\Omega _{\alpha\beta} \Omega ^{\beta \gamma}= \delta ^\gamma_\alpha$. The lower case Latin indices correspond to 5-dimensional fundamental representation of $GL(5)$ ($a,\,b,\,c,\,... = 1,\,...,\,5$). The above generators also obey the relations  $R^{\alpha_1 \alpha_2 }= R^{(\alpha_1 \alpha_2) }$, $R^{a_1a_2a_3 \alpha_1 \alpha_2 }= R^{a_1a_2a_3 (\alpha_1 \alpha_2 )}$, $R^{[a_1a_2,\,b]} = 0$ while the indices on all the other generators are antisymmetric and $\Omega_{\alpha_1 \alpha_2 }$ traceless, for example
$R^{\alpha_1 \ldots \alpha_4 } \Omega_{\alpha_1 \alpha_2 }=0$.
The generators $R^{\alpha_1 \alpha_2 }$ are the generators of Usp(8) which   together with the generators $R^{\alpha_1 \ldots \alpha_4 }$  form    the algebra   $E_6$. The generators $R^{a \alpha_1 \alpha_2 }$ and $  R^{a_1a_2}{}_{\alpha_1 \alpha_2 }$ belong to the 27 and $\overline {27}$-dimensional representations of $E_6$ respectively.
\par
The negative level generators of the $E_{11}$ algebra, down  to level -3, are
 $$ R_{a\alpha_1\alpha_2} ; \ R_{a_1a_2}{}^{\alpha_1\alpha_2} ; \ R_{a_1a_2a_3\alpha_1\alpha_2} , \ R_{a_1a_2a_3\alpha_1\ldots \alpha_4}, \ R_{a_1a_2,b} .
\eqno{(3.3.2)} $$
\par
Up to level 3, the $l_1$ representation decomposes into the following generators
$$
P_a; \ Z_{\alpha_1\alpha_2}; \ Z^a{}_{\alpha_1\alpha_2}; \ Z^{a_1a_2\alpha_1\alpha_2}, \ Z^{a_1a_2\alpha_1\ldots \alpha_4}, \ Z^{ab} \ \eqno{(3.3.3)} $$
where $Z^{a_1a_2\alpha_1\alpha_2} = Z^{a_1a_2(\alpha_1\alpha_2)}$ and $Z^{ab}$ does not have any symmetries on its indices. The commutators of the $E_{11}\otimes_s l_1$ algebra in its five dimensional formulation can be found in [13,15]. The commutators of the $E_{11}$ generators with those of the $l_1$ and $\bar l_1$ representations can be found in appendix A.2.
\par
In the five dimensional theory, the group element $g=g_lg_A$  can be written as
$$
g_l = e^{(x^a P_a +x_{\alpha_1\alpha_2}Z^{\alpha_1\alpha_2} + x_a{}^{\alpha_1\alpha_2} Z^a{}_{\alpha_1\alpha_2} +
  x_{a_1a_2\alpha_1\alpha_2}Z^{a_1a_2\alpha_1\alpha_2} +
 x_{a_1a_2\alpha_1\ldots \alpha_4}Z^{a_1a_2\alpha_1\ldots\alpha_4} + x_{ab}Z^{ab})} \ldots
\eqno{(3.3.4)}$$
$$  g_A =  e^{h_a{}^bK^a{}_b} e^{\varphi_{\alpha_1\alpha_2}R^{\alpha_1\alpha_2} + \varphi_{\alpha_1\ldots \alpha_4}R^{\alpha_1\ldots \alpha_4})}
 e^{A_{a_1a_2}{}^{\alpha_1\alpha_2}R^{a_1a_2}{}_{\alpha_1\alpha_2}}e^{A_{a\alpha_1\alpha_2}R^{a\alpha_1\alpha_2}}\ldots
  \eqno{(3.3.5)} $$
where, $g_l$ is parametrised by
$$x^a ; \ x_{\alpha_1\alpha_2} ; \ x_a{}^{\alpha_1\alpha_2}; \ldots \eqno{(3.3.6)} $$
which we interpret as the coordinates of the  space-time.
We have parametrised $g_A$ by
$$h_a{}^b, \ \varphi_{\alpha_1\alpha_2}, \ \varphi_{\alpha_1\ldots\alpha_4} ;\ A_{a\alpha_1\alpha_2} ; \ A_{a_1a_2}{}^{\alpha_1\alpha_2} ;\ldots . \eqno{(3.3.7)} $$
which are the fields living on the  space-time and we have set the coefficients of the negative level generators in the group element to zero using the local symmetry $I_c(E_{11})$.
 \par
The tangent space  has the components
$$ T^A = (T^a, T_{\alpha_1\alpha_2}, T_{a\alpha_1\alpha_2}, \ldots) . \eqno{(3.3.8)} $$
The level zero transformations of $I_c(E_{11})$ are local Lorentz and Usp(8). At the next levels, the infinitesimal transformation $h \in I_c(E_{11})$ is:
 
$$ h = 1 - \Lambda_{a\alpha_1\alpha_2}(R^{a\alpha_1\alpha_2} - \eta^{ab}\Omega^{\alpha_1\beta_1}\Omega^{\alpha_2\beta_2}R_{b\beta_1\beta_2} ) . \eqno{(3.3.9)} $$
Similarly to previous sections, we can construct any generator of $I_c(E_{11})$ by taking multiple commutators of the generator in this equation. Therefore, if a theory is invariant under these  transformations,  it is invariant under all $I_c(E_{11})$ transformations.
\par
Substituting  the $I_c(E_{11})$ transformation of equation (3.3.9) into equation (2.2) and using the commutators of $E_{11}$ generators and the $l_1$ generators given in Appendix A.2, we find that
 
$$ \delta T^a = -2\Lambda^{a\alpha_1\alpha_2}T_{\alpha_1\alpha_2} , $$
$$ \delta T_{\alpha_1\alpha_2} = \Lambda_{a\alpha_1\alpha_2}T^a + 4\Lambda^a{}_{[\alpha_1 |}{}^{\gamma}T_{a |\alpha_2]\gamma} - {1\over 2}\Omega_{\alpha_1\alpha_2}\Lambda^{a\beta_1\beta_2} T_{a\beta_1\beta_2} , $$
$$ \delta T_{a\alpha_1\alpha_2} = - 4\Lambda_{a[\alpha_1}{}^{\gamma}T_{\alpha_2]\gamma} + {1\over 2} \Omega_{\alpha_1\alpha_2}\Lambda_a{}^{\beta_1\beta_2}T_{\beta_1\beta_2} .  \eqno{(3.3.10)} $$
\par
The  Cartan involution subalgebra invariant tangent space metric for the five dimensional theory is constructed by writing the most general expression quadratic in the tangent space quantities and examining if it is invariant under the transformations of equation (3.3.10).
 We find that the invariant tangent space  element  is given by
$$ \Delta= K_{AB}V^AV^B =
T^aT_a
+ T^{\alpha_1\alpha_2}T_{\alpha_1\alpha_2}
- 2T^{a\alpha_1\alpha_2}T_{a\alpha_1\alpha_2} +\ldots\eqno{(3.3.11)}$$
and reading off the metric we find it is given by
 
$$K_{AB} = \pmatrix{
\eta_{ab} & 0 & 0 \cr
0 & 2(\Omega^{\alpha_1[\beta_1}\Omega^{|\alpha_2|\beta_2]} - {1\over 8}\Omega^{\alpha_1\alpha_2}\Omega^{\beta_1\beta_2}) & 0 \cr
0 & 0 & +2 (\Omega^{\alpha_1[\beta_1}\Omega^{|\alpha_2|\beta_2]} - {1\over 8}\Omega^{\alpha_1\alpha_2}\Omega^{\beta_1\beta_2}) \cr } \eqno{(3.3.12)} $$
\par
We note that had we used the method of  section two which showed that an invariant tangent space metric exists to all level then we would have required the $\bar l_1$ representation. The commutators of these generators with those of $E_{11}$ are given in appendix A.2 for those that want to derive the result using  this method.

\medskip
{\bf 3.4 $E_{11}$ in 4 dimensions}
\medskip
The four dimensional theory has been discussed in [16]. We obtain the four dimensional theory by deleting node 4 from the Dynkin diagram of $E_{11}$:
$$ \matrix{ & & & & & & & & & & & & & & & \bullet & 11 & & & \cr
& & & & & & & & & & & & & & & | & & & & \cr
& \bullet & - & \bullet & - & \bullet & - & \otimes & - & \bullet & - & \bullet & - & \bullet & - & \bullet & - & \bullet & - & \bullet \cr
& 1 & & 2 & & 3 & & 4 & & 5 & & 6 & & 7 & & 8 & & 9 & & 10 \cr} $$
and decomposing the $E_{11} \otimes_s l_1$ algebra into representations of the remaining algebra which is the $GL(4)\times E_7$ algebra.   Up to level 3, the non-negative level generators of the $E_{11}$ algebra, decomposed into the $GL(4) \otimes SL(8)$ representation, are given by
$$ K^a{}_b, \ R^I{}_J; \ R^{I_1\ldots I_4} ; \ R^{aI_1I_2}, \ R^a{}_{I_1I_2} ; \ \hat{K}^{ab} ; \ R^{a_1a_2I}{}_J , \ R^{a_1a_2I_1\ldots I_4}, \ldots . \
\eqno{(3.4.1)} $$
The negative level generators are given by
$$ R_{aI_1I_2}; \ R_a{}^{I_1I_2}; \ R_{a_1a_2}{}^I{}_J, \ R_{a_1a_2I_1\ldots I_4} ,\ldots . \eqno{(3.4.2)} $$
The $l_1$ representation decomposes into the following generators
$$P_a; \ Z^{I_1I_2}, \ Z_{I_1I_2} ; \ Z^a, \ Z^{a}{}^{I}{}_J, \ Z^{aI_1\ldots I_4} , \ldots
\eqno{(3.4.3)} $$
\par
We will be working with the tangent group $I_c(E_{11})$ which at level zero,  is $SO(1,3)\otimes SU(8)=  I_c(GL(4)\times E_7)$  and so we further decompose the $E_{11}$ algebra into this subalgebra. The generators of  $I_c(E_{11})$  are given by
$$
J^a{}_b=K^a{}_b-K^b{}_a , \ J^I{}_J= K^I{}_J-K^J{}_I,\ 
S^{I_1\ldots I_4}= R^{I_1\ldots I_4}-\star R^{I_1\ldots I_4} ,\
$$
$$
S^{aI_1I_2}_{\pm}= R^{aI_1I_2}- \tilde R_{aI_1I_2}
\pm i (R^{a}{}_{I_1I_2}+ \tilde R_{a}{}^{I_1I_2})
$$
$$
S^{a_{1}a_{2}I}{}_{J}=
R^{a_{1}a_{2}I}{}_{J}-\tilde{R}_{a_{1}a_{2}}{}^{J}{}_I,\quad
S^{a_{1}a_{2}I_{1}\cdots I_{4}}= R^{a_{1}a_{2}I_{1}\cdots I_{4}}+\star
\tilde{R}_{a_{1}a_{2}}{}^{I_{1}\cdots I_{4}}, \quad
S^{ab}= \hat K^{ab}-\tilde{\hat K}_{ab} , \ldots
\eqno(3.4.4)$$
The remaining generators in $E_{11}$ are
$$
T^a{}_b=K^a{}_b+K^b{}_a , \ T^I{}_J= K^I{}_J+K^J{}_I,\ 
T^{I_1\ldots I_4}= R^{I_1\ldots I_4}+\star R^{I_1\ldots I_4} ,\
$$
$$
T^{aI_1I_2}_{\pm}= R^{aI_1I_2}+\tilde R_{aI_1I_2}
\pm i (R^{a}{}_{I_1I_2}- \tilde R_{a}{}^{I_1I_2})
$$
$$
T^{a_{1}a_{2}I}{}_{J}=
R^{a_{1}a_{2}I}{}_{J}+\tilde{R}_{a_{1}a_{2}}{}^{J}{}_I,\quad
T^{a_{1}a_{2}I_{1}\cdots I_{4}}= R^{a_{1}a_{2}I_{1}\cdots I_{4}}-\star
\tilde{R}_{a_{1}a_{2}}{}^{I_{1}\cdots I_{4}}, \quad
S^{ab}= \hat K^{ab}+\tilde{\hat K}_{ab} , \ldots
\eqno(3.4.5)$$
The $E_{11}$ algebra expressed in terms of the above generators can be found in reference [16].
\par
The $l_1$ representation when decomposed into representation of $SO(1,3)\otimes SU(8)$ is given by
$$
P_a, \ Z_{\pm}{}^{I_1I_2}, Z^a,\  Z_{+}^{I_1\ldots I_4}, \ Z_{-}^{I_1\ldots I_4}, \ Z_A^a{}^I{}_J , , \ Z_S^a{}^I{}_J , \ldots
\eqno(3.4.6)$$
In terms of the objects of equation (3.4.3) the above are defined by
$$
Z_{\pm}{}^{I_1I_2} = Z^{I_1I_2} \pm iZ_{I_1I_2} ,\
 Z_{+}^{a I_1\ldots I_4}=  Z^{a I_1\ldots I_4}+{1\over 4!} \epsilon ^{I_1\ldots I_4 }{}_{K_1\ldots K_4}  Z^{a K_1\ldots K_4} ,\
$$
$$
Z_{-}^{a I_1\ldots I_4}=  Z^{a I_1\ldots I_4}-{1\over 4!} \epsilon ^{I_1\ldots I_4 }{}_{K_1\ldots K_4}  Z^{a K_1\ldots K_4} ,\
Z_A^a{}^I{}_J = {1\over 2} (Z^a{}^I{}_J -Z^a{}^J{}_I) ,\
Z_S^a{}^I{}_J = {1\over 2} (Z^a{}^I{}_J +Z^a{}^J{}_I)
. \eqno{(3.4.7)} $$
\par
Using appendix A.3, the commutators of the $I_c(E_{11})$ generators $S^{aI_1I_2}_{\pm}$ with the generators of the $L_1$ representation can be found to be given by
$$
[ S_\pm^a{}^{I_1I_2}, P_b]= \delta _b^a Z_\pm{}^{I_1I_2},\quad
[ S_\pm^a{}^{I_1I_2}, Z_\pm{}_{J_1J_2}]=-Z^{aI_1I_2}_{+}{}_{J_1J_2}
\pm 2i\delta^{[I_1}_{[J_1} Z^{a}_S {}_{J_2]}{}^{I_2 ]}
,\quad
$$
$$
[ S_\pm^a{}^{I_1I_2}, Z_\mp{}_{J_1J_2}]=-4  \delta _{J_1J_2}^{I_1I_2}P_a
\pm 2i\delta ^{I_1I_2}_{J_1J_2}Z^a-Z^{aI_1I_2}_{-}{}_{J_1J_2} \pm 2i\delta^{[I_1}_{[J_1} Z^{a}_A{}_{J_2]}{}^{I_2 ]}
$$
$$
[ S_\pm^a{}^{I_1I_2}, Z^b ]= \mp 2i \delta^b_a Z_{\pm}{}^{I_1I_2} ,\
[ S_\pm{}_a{}^{I_1I_2}, Z^b_S{}_J{}_K ]= \pm 8i \delta_a^b
\delta ^{[I_1} _{[J} Z_{\mp}{}^{I_2]} {}_{K]}
$$
$$
[ S_\pm{}_a{}^{I_1I_2}, Z^b_A{}_J{}_K ]= \mp 8i \delta_a^b
\delta ^{[I_1} _{[J} Z_{\pm}{}^{I_2]} {}_{K]},\
$$
$$
[ S_\pm^a{}^{I_1I_2}, Z_+^{b J_1\ldots J_4} ]=
12\delta _a^b( \delta^{I_1I_2}_{[J_1J_2} Z_{\mp J_3J_4]}
+{1\over 4!}\epsilon^{J_1\ldots J_4 I_1I_2 K_1K_2 }Z_{\mp K_1K_2}),
$$
$$
[ S_\pm^a{}^{I_1I_2}, Z_-^{b J_1\ldots J_4} ]=
12\delta _a^b( \delta^{I_1I_2}_{[J_1J_2} Z_{\pm J_3J_4]}
-{1\over 4!}\epsilon^{J_1\ldots J_4 I_1I_2 K_1K_1 }Z_{\pm K_1K_2}),
\eqno(3.4.8)$$
\par
The commutators of the $E_{11} \otimes_s l_1$ algebra when decomposed into representations of $GL(4)\otimes SU(8)$  can be found in reference [16], while  the commutators of the generators of $E_{11}$ with the $l_1$ and $\bar l_1$ generators are given in appendix A.3.
\par
In the four dimensional theory, the group element $g=g_lg_A$ may be written as
$$
g_l = \exp(x^a P_a +x_{\pm I_1I_2}Z^{\pm}{}{}^{I_1I_2} + \hat{x}_aZ^a +
$$
$$
x^{+}_{I_1\ldots I_4} Z_{+}^{I_1\ldots I_4}+x^{+}_{I_1\ldots I_4} Z_{+}^{I_1\ldots I_4}+ x^{-}_{I_1\ldots I_4} Z_{-}^{I_1\ldots I_4}, + x^A_a{}_I{}^J  Z_A^a{}^I{}_J +  x^S_a{}_I{}^J  Z_S^a{}^I{}_J+
\ldots )
\eqno{(3.4.9)} $$
$$ \eqalign{ g_A = & \exp{(h_a{}^bK^a{}_b)}\exp{(\varphi_I{}^JR^I{}_J)}\exp{(\varphi_{I_1\ldots I_4}R^{I_1\ldots I_4})}\exp{(\hat{h}_{ab}\hat{K}^{(ab)})} \times \cr
& \exp{(A_{a_1a_2}{}^J{}_I R^{a_1a_2I}{}_J)}\exp{(A_{a_1a_2I_1\ldots I_4}R^{a_1a_2I_1\ldots I_4})} \exp{(A_{aI_1I_2}R^{aI_2I_2} + A_a{}^{I_1I_2}R^a{}_{I_1I_2})} \cr}
\eqno{(3.4.10)} $$
We have parametrised $g_l$ by
$$ x^a ; \ x_{\pm I_1I_2} ;\ \hat{x}_a ,\ x^{-}_{I_1\ldots I_4},\
 x^{+}_{I_1\ldots I_4},\ x^S_a{}_I{}^J  ,\ x^A_a{}_I{}^J
\ldots
\eqno{(3.4.11)} $$
which we interpret as the coordinates of the generalised space-time.
While we parametrised $g_A$ by the quantities
$$
h_a{}^b, \ \varphi_I{}^J, \ ;\ A_{aI_1I_2}, \ A_a{}^{I_1I_2} ; \ \hat{h}_{ab} , \ A_{a_1a_2}{}^I{}_J, \ A_{a_1a_2I_1\ldots I_4} .
\eqno{(3.4.12)} $$
which are the fields of the theory living on the generalised space-time. In writing the group element we have used the local $I_c(E_{11})$ symmetry to set to zero all the coefficients of the negative level generators in the group element.
\par
We may raise and lower the lower case latin (Lorentz) indices with $\delta^{ab}$, and similarly we raise and lower the capital latin (internal) indices with $\delta^{IJ}$ and their inverses respectively.
\par
The tangent object in four dimensions contains the components
$$
V^A = (V^a, V_{\pm I_1I_2}, \hat{V}_a, \ V^{+}_{I_1\ldots I_4} ,\
\ V^{-}_{I_1\ldots I_4}  ,\ V^S_a{}_I{}^J ,\ V^A_a{}_I{}^J ,\
\ldots )
. \eqno{(3.4.13)} $$
The level zero transformations of $I_c(E_{11})$ are local Lorentz and SU(8)  transformations. In four dimensions the infinitesimal transformation $h \in I_c(E_{11})$ at the next levels are of the form
$$
h = 1 - \sum_{\pm}\Lambda_{a\pm I_1I_2}S_{\pm}^{aI_1I_2}
\eqno{(3.4.14)} $$
We may construct any generator of $I_c(E_{11})$ by simply taking multiple commutators of the generator that occurs in this equation.
\par
Substituting the above $I_c(E_{11})$ transformation into equation (2.2) and using the commutators of $E_{11}$ generators with the $l_1$ generators, given in Appendix A.3, we find that
$$
\delta V^a = -4\sum_{\pm}\Lambda_{a\pm J_1J_2}V_{\mp}{}^{ I_1I_2} $$
$$ \delta V_{\pm I_1I_2} = \Lambda_{a\pm I_1I_2}V^a \mp 2i\Lambda_{a \pm I_1I_2}\hat{V}^a +24 \Lambda_{a\mp J_1J_2}V_a^{+}{}^{J_1J_2}{}_{I_1I_2}
+24 \Lambda_{a\pm J_1J_2}V_a^{-}{}^{J_1J_2}{}_{I_1I_2}
$$
$$
\mp 8i \Lambda_{a\pm J [ I_1}V^A_a{}^{J}{}_{|I_2]}
\mp 8i \Lambda_{a\mp J [ I_1}V^S_a{}^{J}{}_{|I_2]}
$$

$$
\delta \hat{V}_a = 2i\sum_{\pm}\pm\Lambda_{a\pm I_1I_2}V_{\mp}{}^{ I_1I_2}
$$
$$
 \delta V^{+}_a{}_{I_1\ldots I_4} = -{1\over 2} \sum_{\pm} (\Lambda_{a\pm [ I_1I_2} V_{\pm I_3I_4]}+{1\over 4!}\epsilon _{I_1\ldots I_4 }{}^{K_1\ldots K_4}
\Lambda_{a\pm  K_1K_2} V_{\pm K_3K_4} ),
$$
$$
\delta V^{-}_a{}_{I_1\ldots I_4} =-{1\over 2} \sum_{\pm} (\Lambda_{a\pm [ I_1I_2}
V_{\mp I_3I_4]} -{1\over 4!}\epsilon _{I_1\ldots I_4 }{}^{K_1\ldots K_4}
\Lambda_{a\pm  K_1K_2} V_{\mp K_3K_4} )
$$
$$
\delta V_S^a {}_I{}_J= -\sum_{\pm} \pm 2i \Lambda _{a\pm L(J} V_{\pm I) }{}^L ,\ \quad
\delta V_A^a {}_I{}_J=-\sum_{\pm} \pm 2i \Lambda _{a\pm L[J} V_{\mp I] }{}^L
\eqno{(3.4.15)}$$
\par
The Cartan involution subalgebra, $I_c(E_{11})$,  invariant tangent space metric for the four dimensional theory is constructed by writing the most general expression quadratic in the tangent space quantities and examining if it is invariant under the transformations of equation (3.4.10).
We find that the invariant quantity is given by
$$
\Delta = K_{AB}V^AV^B =V^aV_a
+ 4\sum_{\pm}V_{\pm I_1I_2}V_{\mp}{}^{ I_1I_2}
+ 4 \hat{V}_a\hat{V}^a +24V^{+}_{I_1\ldots I_4} V^{+}{}^{I_1\ldots I_4}
$$
$$
+24V^{-}_{I_1\ldots I_4} V^{-}{}^{I_1\ldots I_4}
+4 V^S_a{}^I{}_J V^S{}^{a}{}_I{}^J +4 V^A_a{}^I{}_J V^A{}^{a}{}_I {}^J +\ldots
\eqno{(3.4.16)} $$
Reading off the invariant metric we find that it is given, up  to the levels considered, by 
\hoffset-1.5cm
\def\quad{\hskip1ex\relax}
$$
K_{AB} = \pmatrix{ \delta_{ab} & 0 & 0 & 0 & 0 & 0 & 0 & 0 \cr
0 & 0 & 4\delta^{I_1I_2, J_1J_2} & 0 & 0 & 0 & 0 & 0 \cr
0 & 4\delta^{I_1I_2, J_1J_2} & 0 & 0 & 0 & 0 & 0 & 0 \cr
0 & 0 & 0 & 4\delta^{ab} & 0 & 0 & 0 & 0 \cr
0 & 0 & 0 & 0 & 24 \delta^{I_1\ldots I_4, J_1\ldots J_4} & 0 & 0 & 0\cr
0 & 0 & 0 & 0 & 0   &24 \delta^{I_1\ldots I_4, J_1\ldots J_4} & 0 & 0 \cr
0 & 0 & 0 & 0 & 0 &0 & 4 \delta^{ab}\delta_{IK}\delta^{JL} & 0 \cr
0 & 0 & 0 & 0 & 0 &0 & 0 & 4 \delta^{ab}\delta_{IK}\delta^{JL} \cr}
\eqno{(3.4.17)}$$
\hoffset-0.2cm

\medskip
{\bf 3.5 $A_1^{+++}$}
\medskip

It has  been  conjectured that the non-linear realisation of the semi-direct product  of $A_1^{+++}$ with its first fundamental, or vector,  representation, denoted by $A_1^{+++}\otimes_s l_1 $, leads to  the complete low energy effective action for four dimensional gravity [17]. Here we have denoted the very extension of  $A_1$ by $A_1^{+++}$.
The four dimensional theory appears when we delete node 4 in the Dynkin diagram,
$$ \matrix{ \bullet & - & \bullet & - &\bullet & = & \otimes \cr
1 & & 2 & & 3 & & 4 \cr} $$
and decompose the algebra into the remaining algebra which is  $GL(4)$. Up to level 2, the non-negative level generators of the decomposed $A_1^{+++}$ algebra are given by [18]
$$ K^a{}_b; R^{ab}; \ R^{ab,cd} , \eqno{(3.5.1)} $$
which obey $R^{ab} = R^{(ab)}$ and $R^{ab,cd} = R^{ab,(cd)} = R^{[ab],cd}$, while  the negative level generators are given by
$$ R_{ab} ; \ R_{ab,cd} . \eqno{(3.5.2)} $$
\par
Up to level 2, the $l_1$ representation contains the following generators
$$ P_a ; \ Z^a ; \ Z^{abc}, Z^{ab,c}, \eqno{(3.5.3)} \ $$
where $Z^{abc} = Z^{(abc)}$, $Z^{ab,c} = Z^{[ab],c}$ and $Z^{[ab,c]} = 0$.
The commutators of the $A_1^{+++}\otimes_s l_1$ algebra is given in reference [18].
\par
The group element $g=g_lg_A$ can be written as
$$ g_l = \exp{(x^aP_a + y_a Z^a + x_{abc}Z^{abc} + x_{ab,c}Z^{ab,c})}, $$
$$ g_A = \exp{(h_a{}^bK^a{}_b)}\exp{(A_{ab,cd}R^{ab,cd})}\exp{(A_{ab}R^{ab})} . \eqno{(3.5.4)} $$
in which $g_l$ is parametrised by the  space-time coordinates
$$ x^a; \ y_a; \ x_{abc}, \ x_{ab,c}, \eqno{(3.5.5)} $$
and $g_A$ is parametrised by the fields
$$ h_a{}^b; A_{ab}; A_{ab,cd} , \eqno{(3.5.6)} $$
which live on the  space-time and have the same symmetries as their corresponding generators. The coefficients of the negative level generators in $g_A$ are set to zero using the local symmetry $I_c(A_1^{+++})$. The field $h_a{}^b$ is the graviton, while $A_{ab}$ is the dual graviton.
\par
The components of the tangent object is given by
$$V^A = (T^a, \hat{T}_a, T_{abc}, T_{ab,c}, \ldots) . \eqno{(3.5.7)} $$
The level zero transformations of $I_c(E_{11})$ are SO(4). At the next levels, the infinitesimal transformation $h \in I_c(A_1^{+++})$ is of the form
$$
h = 1 - \Lambda_{ab}(R^{ab} - \eta^{ab}\eta^{cd}R_{cd}) .
\eqno{(3.5.8)}$$
 Again, we may construct any generator of $I_c(A_1^{+++})$ by taking multiple commutators of the generators in the above equation.
\par
Substituting the  $I_c(A_1^{+++})$ transformation of equation (3.5.8) into equation (2.2) and using the commutators given in Appendix A5, we find
$$ \delta T^d = - 2\Lambda^{cd}\hat{T}_c ,\quad \delta \hat{T}_d = \Lambda_{cd}T^c - 2\Lambda^{ce}T_{ced} - {8\over 3}\Lambda^{ce}T_{dc,e} , $$
$$ \delta T_{abc} = \Lambda_{(ab}\hat{T}_{c)} , \quad \delta T_{ab,c} = \Lambda_{c[a}\hat{T}_{b]} . \eqno{(3.5.9)} $$
where indices are raised and lowered with the Kronecker delta.

The tangent space metric which is invariant under the Cartan involution invariant subalgebra is found to be given by
$$ K_{AB}V^AV^B = T^aT_a + 2\hat{T}_a\hat{T}^a + 4T_{abc}T^{abc} + {16\over3}T_{ab,c}T^{ab,c} +\dots ,
\eqno{(3.5.10)} $$
and we find that the invariant metric, up to the levels considered,  is given by
$$
K_{AB} = \pmatrix{ \delta_{ab} & 0 & 0 & 0 \cr
0 & 2\delta^{ab} & 0 & 0 \cr
0 & 0 & 4\delta^{abc,def} & 0 \cr
0 & 0 & 0 & {16\over3}(\delta^{ab,de}\delta^{cf} - \delta^{abc,def}) \cr} \eqno{(3.5.11)} $$


\medskip
{\bf 4. The gauge fixed multiplet}
\medskip
In this section we use the  invariant metric discussed above to construct an invariant set of $E_{11}\otimes_s l_1$ equations. We will see that these can be interpreted as gauge fixing conditions. The $E_{11}$ Cartan forms are defined in equation (1.8).   When the $R^{\underline \alpha}$ generators are taken  to belong to the $l_1$ representation and the first index, which is  associated with the $l_1$ representation, is taken to be a tangent index the Cartan form can be written as [19]
$$
G_{A, B}{}^{C}\equiv E_A{}^{\Pi} G_{\Pi, B}{}^{C}= E_A{}^{\Pi} G_{\Pi, \underline \alpha} (D^{\underline \alpha})_{B}{}^{C}= E_A{}^{\Pi} E_B{}^{\Lambda}\partial_{\Pi} E_{\Lambda}{}^C .
\eqno(4.1)$$
This quantitity is invariant under the rigid $E_{11}\otimes_s l_1$ transformations of equation (1.4) but it transforms under the local $I_c(E_{11})$ transformations as its indices indicate.
\par
Equipped with the invariant tangent metric we can consider the quantity
$$ 
G^C\equiv K^{AB} G_{A, B}{}^{C} ,
\eqno(4.2)$$
which transforms under the local $I_c(E_{11})$ transformations on its remaining uncontracted index $C$.  Indeed it transforms like the tangent quantity $V^A$ of equation (2.2). Clearly,  we can set
$$G^C=0 ,
\eqno(4.3)$$
 and still preserve the symmetries of the non-linear realisation.
\par
At the linearised level
$$
G_{A, B}{}^{C}= \partial_{A} E_{B}{}^C= \partial_{A}  A_{\underline \alpha}(D^{\underline \alpha})_{B}{}^C ,
\eqno(4.4)$$
since $E_{\Pi}{}^A= (e^{{\cal A}})_{\Pi}{}^A$ where ${\cal A}_{B}{}^C= A_{\underline \alpha}(D^{\underline \alpha})_{B}{}^C$. As a result
$$G^C=K^{AB}\partial_{A}  A_{\underline \alpha}(D^{\underline \alpha})_{B}{}^C= \eta^{ab} \partial_{a}  A_{\underline \alpha}(D^{\underline \alpha})_{b}{}^C+ \ldots ,
\eqno(4.5)$$
where $+ \ldots $ means terms involving higher level derivatives.
We recall that
$$
[R^{\underline \alpha} ,P_b]= - (D^{\underline \alpha})_b{}^Cl_C ,
\eqno(4.6)$$
and so we can read off this matrix from the algebra and find the linearised result in a simple way.
We now evaluate this quantity $G^C$ for the $E_{11}\otimes_s l_1$ in  eleven, five and four dimensional non-linear realisation.
\medskip
{\bf 4.1 Eleven dimensions}
\medskip
We will now evaluate the quantity of equation (4.2) in the eleven dimensional theory in terms of the fields of equation (3.2.10). This depends on the generalised vielbein $E_\Pi{}^A$ which  was given in terms of these fields, up to level three, in equation (4.3) of reference [18]. We will denote the  generalised space-time derivatives with respect to the coordinates of equation (3.2.9) by
$$
\partial_{\Pi}= \{
 \partial_a ; \ \partial^{a_1a_2} ; \ \partial^{a_1\ldots a_5} ; \ \partial^{a_1\ldots a_8} , \ \partial^{a_1\ldots a_7, b} , \ldots \} . \eqno(4.1.1) $$
\par
Substituting  the generalised vielbein into the Cartan form of equation (4.1) and contract ing the resulting Cartan form with the metric $K^{AB}$ of equation (3.2.18) we find the object $G^C$ has the components
 $$
 K^A{}^BG_{A,}{}_B{}^{c} = (\det{e})^{1\over2} (\partial^ah_a{}^c - {1\over2}\partial^ch_a{}^a ) ,
\eqno{(4.1.2)} $$
$$ K^A{}^BG_{A,}{}_B{}_{c_1c_2} = (\det{e})^{1\over2}\big(-3\partial^aA_{ac_1c_2} -\partial_{[c_1|a|}h_{c_2]}{}^{a} - {1\over4}\partial_{c_1c_2}h_a{}^a\big) ,
\eqno{(4.1.3)}$$
$$ K^A{}^BG_{A,}{}_B{}_{c_1\ldots c_5} = -(\det{e})^{1\over2}\big(3\partial^aA_{ac_1\ldots c_5} - {1\over2}\partial_{[c_1c_2}A_{c_3c_4c_5]} - {1\over24}\partial_{[c_1\ldots c_4|a|}h_{c_5]}{}^{a} - {1\over240}\partial_{c_1\ldots c_5}h_a{}^a\big) ,
\eqno{(4.1.4)} $$
$$ K^A{}^BG_{A,}{}_B{}_{c_1\ldots c_8} = -(\det{e})^{1\over2}\big({3\over2}\partial^aA_{c_1\ldots c_8,a} + {1\over2}\partial_{[c_1c_2}A_{c_3\ldots c_8]} + {1\over120}\partial_{[c_1\ldots c_5}A_{c_6c_7c_8]} $$
$$ + {1\over630}(\partial_{[c_1\ldots c_7|a|}h_{c_8]}{}^{a} + {1\over{120 \times 84}}\partial_{c_1\ldots c_8}h_d{}^d)\big) ,
\eqno{(4.1.5)} $$
 $$ K^A{}^BG_{A,}{}_B{}_{c_1\ldots c_7,c} = -(\det{e})^{1\over2}(\partial^{a}({4\over3}A_{a[c_1\ldots c_7,c]} - {4\over3}A_{a[c_1\ldots c_7],c}) +{1\over2}\partial_{[c_1c_2}(-A_{c_3\ldots c_7c]} + A_{c_3\ldots c_7]c}) $$
$$ -{1\over 120}\partial_{[c_1\ldots c_5}(A_{c_6c_7c]}-A_{c_6c_7]c})  +{1\over{30\times945}}(7\partial_{[c_1\ldots c_6|a,c|}h_{c_7]}{}^{a} - 8\partial_{[c_1\ldots c_7,|a|}h_{c]}{}^{a}+ {1\over2}\partial_{c_1\ldots c_7,c}h_d{}^d)\big) , \eqno{(4.1.6)} $$
where $\delta_{a_1\ldots a_n, b_1\ldots b_n} = \delta_{[a_1|b_1|}\delta_{a_2|b_2|}\ldots \delta_{a_n]b_n}$
\par
As mentioned above we can  set $G^C=0$
and preserve the symmetries of the non-linear realisation. Doing  this we recognise the first component of equation (4.1.2)  as the De Donder gauge fixing condition for gravity. The next equation  is  a gauge fixing condition for the three form field and similarly for higher components. Indeed the entire multiplet can be viewed as a gauge fixing condition for the gauge transformations given for the fields of the $E_{11}\otimes_s l_1$ non-linear realisation in reference [19]. As such we have found that we can fix the gauge symmetries in an $E_{11}$ covariant manner in eleven dimensions.


\medskip
{\bf 4.2 Five dimensions}
\medskip
We now find the quantity $G^C$, given in equation (4.2),  in the five dimensional theory in terms of the fields of equation (3.3.7), using the generalised vielbein $E_{\Pi}{}^A$, which was given in terms of these fields, up to level one, in equation (2.14) of reference [18]. We denote the generalised space-time derivatives with respect to the coordinates of equation (3.3.6) by
$$
\partial_{\Pi}= \{
\partial_a ; \ \partial^{\alpha_1\alpha_2} ; \ \partial^a{}_{\alpha_1\alpha_2}, \ldots \} . \eqno(4.2.1) $$
\par
We substitute the generalised vielbein into the Cartan form of equation (4.1) and contract the resulting Cartan form with the metric $K^{AB}$ of equation (3.3.12) to find that $G^C$ has the components
$$ K^A{}^BG_{A,}{}_B{}^{c} = (\det{e})^{1\over2}(\partial^ah_a{}^c - {1\over 2}\partial^ch_a{}^a) ,
\eqno(4.2.2) $$
$$ K^A{}^BG_{A,}{}_{B\kappa_1\kappa_2} = -(\det{e})^{1\over2}(\partial^aA_{a\kappa_1\kappa_2} -{1\over2}(f^{-1})_{\nu_1\nu_2}^{\alpha_1\alpha_2}\partial_{\alpha_1\alpha_2}f^{\nu_1\nu_2}_{\kappa_1\kappa_2} + {1\over2}\partial_{\kappa_1\kappa_2}h_a{}^a) ,
\eqno(4.2.3) $$
$$ K^{AB}G_{A,B}{}_{c\kappa_1\kappa_2} = -(\det{e})^{1\over2}(2\partial^aA_{a\kappa_1\kappa_2} + 2\Omega^{[\gamma_1[\alpha_1}\delta^{\gamma_2]\alpha_2]}_{\kappa_1\kappa_2}\partial_{\alpha_1\alpha_2}A_{a\gamma_1\gamma_2}
$$
$$+ {1\over2}(f^{-1})_{\kappa_1\kappa_2}^{\nu_1\nu_2}\partial_{c\alpha_1\alpha_2}(f^{\alpha_1\alpha_2}_{\nu_1\nu_2}) - \partial_{a\kappa_1\kappa_2}h_c{}^a - {1\over2}\partial_{c\kappa_1\kappa_2}h_a{}^a) . \eqno(4.2.4) $$
ÔøΩwhere $f$ is a function of the scalar fields whose definition can be found in equation (3.3.14) of reference [18].
\par
Setting $G^C=0$
preserves the symmetries of the non-linear realisation and can be recognised as a gauge fixing condition.
\medskip
{\bf 4.3 Four dimensions}
\medskip
We will now evaluate the quantity of equation (4.2) in the four dimensional theory in terms of the fields of equation (3.4.7). This depends on the generalised vielbein $E_{\Pi}{}^A$ which was given in terms of these fields, up to level three, in equation (3.4.12) of reference [18]. The generalised space-time derivatives with respect to the coordinates of equation (3.4.6) are
$$
\partial_{\Pi}= \{
\partial_a ; \ \partial_{\pm}^{I_1I_2} ; \ \hat\partial^{{a}}, \ldots \} . \eqno(4.3.1) $$
\par
Substituting the generalised vielbein into the Cartan form of equation (4.1) and contract the resulting Cartan form with the metric $K^{AB}$ of equation (3.4.12) and we find $G^C$ has the components


$$ G^c = (\det{e})^{{1 \over 2}} (\partial^ah_a{}^c - {1\over 2}\partial^ch_a{}^a) , $$

$$ G_{+K_1K_2} = (\det{e})^{1\over2}(-{1\over2}\partial^aA_{aK_1K_2} - {1\over2}\partial^aA_a{}^{K_1K_2} - {1\over8}\partial_{-K_1K_2}h_a{}^a $$
$$ + {1\over4}(\partial_{-I[K_1}\phi_{K_2]}{}^I - \partial^{-I[K_1}\phi_I{}^{K_2]} + {1\over4}\partial_{-K_1K_2}\phi_I{}^I) + {1\over4}(\partial_{+I[K_1}\phi_{K_2]}{}^I + \partial^{+I[K_1}\phi_I{}^{K_2]}) $$
$$ - {1\over4}({i\over2}\partial^{-I_1I_2}\phi_{I_1I_2K_1K_2} + {i\over 2(4!)} \partial^{-I_1I_2}*\phi_{I_1I_2K_1K_2})$$
$$ + {1\over4}({i\over2}\partial^{+I_1I_2}\phi_{I_1I_2K_1K_2} + {i\over 2(4!)} \partial^{+I_1I_2}*\phi_{I_1I_2K_1K_2})) , $$

$$  G_{-K_1K_2} = (\det{e})^{1\over2}(-{1\over2}\partial^aA_{aK_1K_2} + {1\over2}\partial^aA_a{}^{K_1K_2} - {1\over8}\partial_{+K_1K_2}h_a{}^a $$
$$ + {1\over4}(\partial_{+I[K_1}\phi_{K_2]}{}^I - \partial^{+I[K_1}\phi_I{}^{K_2]} + {1\over4}\partial_{+K_1K_2}\phi_I{}^I) + {1\over4}(\partial_{-I[K_1}\phi_{K_2]}{}^I + \partial^{-I[K_1}\phi_I{}^{K_2]}) $$
$$ + {1\over4}({i\over2}\partial^{+I_1I_2}\phi_{I_1I_2K_1K_2} + {i\over 2(4!)} \partial^{+I_1I_2}*\phi_{I_1I_2K_1K_2}) $$
$$ + {1\over4}({i\over2}\partial^{-I_1I_2}\phi_{I_1I_2K_1K_2} + {i\over 2(4!)} \partial^{-I_1I_2}*\phi_{I_1I_2K_1K_2})) , $$

$$ G_{c} = -(\det{e})^{1\over2}({1\over2}\partial_ah^{(ac)} + {1\over4}(\partial^ah_a{}^c + {1\over2}\partial_ch_a{}^a) , $$

$$ G^{+cK_1\ldots K_4} = (\det{e})^{1\over2}({1\over 12}\partial^aA_{acK_1\ldots K_4} - {1\over24}(\partial_{+aK_1\ldots K_4}h_a{}^c + {1\over2}\partial_{+cK_1\ldots K_4}h_a{}^a) $$
$$ - {5\over24}\partial_{+c[K_1\ldots K_1}\phi_{I]}{}^I + {1\over4}({5\over12}(\partial^A{}_c{}_I{}^{[I}A^{K_1\ldots K_4]} - \partial^A{}_c{}_{[I}{}^IA_{K_1\ldots K_4]}) +{1\over 8}\partial^A{}_c{}_I{}^{I}A_{K_1\ldots K_4}) $$
$$ + {1\over4}({5\over 12}( \partial^S{}_c{}_I{}^{[I}A^{K_1\ldots K_4]} + \partial^S{}_c{}_{[I}{}^IA_{K_1\ldots K_4]}) + {1\over 3}\partial^S{}_c{}_I{}^{I}A_{K_1\ldots K_4})) , $$

$$ G^{-cK_1\ldots K_4} = (\det{e})^{1\over2}({1\over 12}\partial^aA_{acK_1\ldots K_4} - {1\over24}(\partial_{-aK_1\ldots K_4}h_a{}^c + {1\over2}\partial_{-cK_1\ldots K_4}h_a{}^a) $$
$$ - {5\over24}\partial_{-c[K_1\ldots K_1}\phi_{I]}{}^I + {1\over4}({5\over12}(\partial^A{}_c{}_I{}^{[I}A^{K_1\ldots K_4]} - \partial^A{}_c{}_{[I}{}^IA_{K_1\ldots K_4]}) +{1\over 8}\partial^A{}_c{}_I{}^{I}A_{K_1\ldots K_4}) $$
$$ + {1\over4}({5\over 12}( \partial^S{}_c{}_I{}^{[I}A^{K_1\ldots K_4]} + \partial^S{}_c{}_{[I}{}^IA_{K_1\ldots K_4]}) + {1\over 3}\partial^S{}_c{}_I{}^{I}A_{K_1\ldots K_4})) , $$

$$ G_A{}^{cK_1}{}_{K_2} = (\det{e})^{1\over2}({1\over2}\partial^aA_{acK_1}{}^{K_2} + {2\over(4!)^2}(+{5\over4}\delta^{[K_2}_{K_1}\partial^{+I_1\ldots I_4]}*A_{I_1\ldots I_4} - {1\over4}\delta^{K_2}_{K_1}\partial^{+cI_1\ldots I_4}*A_{I_1\ldots I_4}) $$
$$ + {2\over(4!)^2}(+{5\over4}\delta^{[K_2}_{K_1}\partial^{-I_1\ldots I_4]}*A_{I_1\ldots I_4} - {1\over4}\delta^{K_2}_{K_1}\partial^{-cI_1\ldots I_4}*A_{I_1\ldots I_4}) $$
$$ - {1\over4}(\partial_A{}^{aK_1}{}_{K_2}h_a{}^c + {1\over2}\partial_A{}^{cK_1}{}_{K_2}h_a{}^a) - {1\over 4}(\partial_A{}^{cJ}{}_{K_2}\phi_{K_1}{}^J - \partial_A{}^{cK_1}{}_{J}\phi_J{}^{K_2})) , $$

$$ G_A{}^{cK_1}{}_{K_2} = (\det{e})^{1\over2}({1\over2}\partial^aA_{acK_1}{}^{K_2} + {2\over(4!)^2}(+{5\over4}\delta^{[K_2}_{K_1}\partial^{+I_1\ldots I_4]}*A_{I_1\ldots I_4} - {1\over4}\delta^{K_2}_{K_1}\partial^{+cI_1\ldots I_4}*A_{I_1\ldots I_4}) $$
$$ + {2\over(4!)^2}(+{5\over4}\delta^{[K_2}_{K_1}\partial^{-I_1\ldots I_4]}*A_{I_1\ldots I_4} - {1\over4}\delta^{K_2}_{K_1}\partial^{-cI_1\ldots I_4}*A_{I_1\ldots I_4}) $$
$$ - {1\over4}(\partial_S{}^{aK_1}{}_{K_2}h_a{}^c + {1\over2}\partial_S{}^{cK_1}{}_{K_2}h_a{}^a) - {1\over 4}(\partial_S{}^{cJ}{}_{K_2}\phi_{K_1}{}^J - \partial_S{}^{cK_1}{}_{J}\phi_J{}^{K_2})) . $$

\par
We may set $G^C=0$ and we recognise the result as a set of gauge fixing conditions.
\medskip
{\bf 4.4 $A_1^{+++}$}
\medskip
In this subsection we  find the quantity $G^C$ of equation (4.2) in terms of  the fields given in equation (3.5.6). The vielbein $E_{\Pi}{}^A$ was given in terms of these fields, up to level two, was found in equation (4.25) of reference [18]. The generalised space-time derivatives with respect to the coordinates of equation (3.5.5) are given by
$$
\partial_{\Pi} = \{
\partial_a ; \ \partial^{\hat{a}} ; \ \partial_{abc}, \ \partial_{ab,c} , \ldots \}. \eqno(4.4.1) $$
\par
We substitute the generalised vielbein into the Cartan form of equation (4.1), and contract the result with the metric of equation (3.5.11) and we find that $G^C$ has the components
 
$$ K^{AB}G_{A,B}{}^c = (\det{e})^{1\over2}(\partial^ah_{a}{}^c - {1\over2}\partial^ch_a{}^a) , \eqno(4.4.2) $$
$$ K^{AB}G_{A,B}{}_{c} = -(\det{e})^{1\over2}(\partial^a A_{ac} + {1\over2}\partial_{\hat{a}}h_c{}^a + {1\over2}\partial_{\hat{c}}h_a{}^a) , \eqno(4.4.3) $$
$$ K^{AB}G_{A,B}{}_{c_1c_2c_3} = (\det{e})^{1\over2}(\partial^a(A_{a(c_1,c_2c_3)}) -{1\over2}\partial_{[\hat{c}_1}A_{c_2c_3]} - {3\over4}\partial_{(c_1c_2|a|}h_{c_3)}{}^{a}) - {1\over8}\partial_{c_1c_2c_3}h_a{}^a) , \eqno(4.4.4) $$
$$ K^{AB}G_{A,B}{}_{[c_1c_2],c_3} = (\det{e})^{1\over2}(\partial^a({3\over4} A_{[c_1c_2],c_3a} - {1\over2}A_{a[c_1,c_2]c_3}) -{1\over2}\partial_{[\hat{c}_1}A_{c_2]c_3} $$
$$ - {3\over16}(2\partial_{[c_1|a,c_3|}h_{c_2]}{}^{a} + \partial_{[c_1c_2],a}h_{c_3}{}^{a} - 3\partial_{[c_1c_2|,a|}h_{c_3]}{}^{a} + {1\over2}\partial_{[c_1c_2],c_3}h_a{}^a)) . \eqno(4.4.5) $$
\par
Again we recognise the equation $G^C = 0$ as a set of gauge fixing conditions.

\medskip
{\bf {Acknowledgements}}
\medskip
We thank Axel Kleinschmidt for discussions.   We also wish to thank the SFTC for support from Consolidated grant number ST/J002798/1 and Michaella Pettit wishes to thanks EPSRC  for the support provided by the  Research Studentship EP/M50788X/1.

\medskip

{\bf Appendix A.1}
\medskip
The commutators of the $E_{11}$ algebra can be found in the book of reference [10]. In this appendix we give the commutators of the $E_{11}$ generators with the $l_1$ generators and the $\bar l_1$ generators.
\par
ÔøΩ
The commutators of $l_1$ generators with $E_{11}$ generators are:
$$ [K^a{}_b,P_c] = -\delta_c^aP_b + {1\over2}\delta^a_bP_c, \ \ \ \ [K^a{}_b,Z^{a_1a_2}] = 2\delta_b^{[a_1}Z^{|a|a_2]} + {1\over2}\delta^a_bZ^{a_1a_2} $$
$$ [K^a{}_b,Z^{a_1\ldots a_5}] = 5\delta_b^{[a_1}Z^{|a|a_2\ldots a_5]} + {1\over2}\delta^a_bZ^{a_1\ldots a_5} $$
$$ [K^a{}_b,Z^{a_1\ldots a_8}] = 8\delta_b^{[a_1}Z^{|a|a_2\ldots a_8]} + {1\over2}\delta^a_bZ^{a_1\ldots a_8} $$
$$ [K^a{}_b,Z^{a_1\ldots a_7,c}] = 7\delta_b^{[a_1}Z^{|a|a_2\ldots a_7],c} +\delta_b^cZ^{a_1\ldots a_7,a} + {1\over2}\delta^a_bZ^{a_1\ldots a_7,c} \eqno(A.1.1) $$
The commutators with the positive level $E_{11}$ generators are:
$$ [R^{a_1a_2a_3}, P_a] = 3\delta_a^{[a_1}Z^{a_2a_3]}, \ \ \ \ [R^{a_1a_2a_3}, Z^{a_4a_5}] = Z^{a_1\ldots a_5} $$
$$[R^{a_1a_2a_3},Z^{b_1\ldots b_5}] = Z^{b_1\ldots b_5a_1a_2a_3} + Z^{b_1\ldots b_5[a_1a_2,a_3]} $$
$$[R^{a_1\ldots a_6},P_a] = -3\delta_a^{[a_1}Z^{a_2\ldots a_6]}, \ \ \ \ [R^{a_1\ldots a_6},Z^{b_1b_2}] = -Z^{b_1b_2a_1\ldots a_6} - Z^{b_1b_2[a_1\ldots a_5,a_6]} $$
$$[R^{a_1\ldots a_8,a}, P_b] = -{4\over3}\delta^a_bZ^{a_1\ldots a_8} + {4\over3}\delta_b^{[a_1}Z^{a_2\ldots a_8]a} + {4\over3}\delta_b^{[a_1}Z^{a_2\ldots a_8],a} \eqno(A.1.2) $$
The commutators with the negative level $E_{11}$ generators are:
$$ [R_{a_1a_2a_3},P_a] = 0, \ \ \ \ [R_{a_1a_2a_3},Z^{b_1b_2}] = 6\delta^{b_1b_2}_{[a_1a_2}P_{a_3]} $$
$$ [R_{a_1a_2a_3},Z^{b_1\ldots b_5}] = 60\delta_{a_1a_2a_3}^{[b_1b_2b_3}Z^{b_4b_5]}, \ \ \ \ [R_{a_1a_2a_3}, Z^{b_1\ldots b_8}] = -42\delta_{a_1a_2a_3}^{[b_1b_2b_3}Z^{b_4\ldots b_8]} $$
$$ [R_{a_1a_2a_3},Z^{b_1\ldots b_7,b}] = {945\over4}\delta_{a_1a_2a_3}^{[b_1b_2b_3}Z^{b_4\ldots b_7]b} + {945\over 4}\delta_{a_1a_2a_3}^{[b_1b_2|b|}Z^{b_3\ldots b_7]} \eqno(A.1.3) $$
\par
Using the action of the Cartan involution of equation (3.2.3), we find that the commutators of the $\bar{l}_1$ generators with the $E_{11}$ generators are given by, at level 0:
$$ [K^a{}_b,\bar{P}^c] = \delta_b^c\bar{P}^a - {1\over2}\delta^a_b\bar{P}^c, \ \ \ \ [K^a{}_b,\bar{Z}_{a_1a_2}] = -2\delta^a_{[a_1}\bar{Z}_{|b|a_2]} - {1\over2}\delta^a_b\bar{Z}_{a_1a_2} $$
$$ [K^a{}_b,\bar{Z}_{a_1\ldots a_5}] = -5\delta^a_{[a_1}\bar{Z}_{|b|a_2\ldots a_5]} - {1\over2}\delta^a_b\bar{Z}_{a_1\ldots a_5} $$
$$ [K^a{}_b,\bar{Z}_{a_1\ldots a_8}] = -8\delta^a_{[a_1}\bar{Z}_{|b|a_2\ldots a_8]} - {1\over2}\delta^a_b\bar{Z}_{a_1\ldots a_8} $$
$$ [K^a{}_b,\bar{Z}_{a_1\ldots a_7,c}] = -7\delta^a_{[a_1}\bar{Z}_{|b|a_2\ldots a_7],c} - \delta^a_c\bar{Z}_{a_1\ldots a_7,b} - {1\over2}\delta^a_b\bar{Z}_{a_1\ldots a_7,c} \eqno(A.1.4) $$
The commutators with the positive level $E_{11}$ generators are:
$$ [R^{a_1a_2a_3}, \bar{P}^a] = 0, \ \ \ \ [R^{a_1a_2a_3}, \bar{Z}_{b_1b_2}] = -6\delta_{b_1b_2}^{[a_1a_2}\bar{P}^{a_3]} $$
$$ [R^{a_1a_2a_3}, \bar{Z}_{b_1 \ldots b_5}] = -60\delta_{[b_1b_2b_3}^{a_1a_2a_3}\bar{Z}_{b_4b_5]}, \ \ \ \ [R^{a_1a_2a_3}, \bar{Z}_{b_1 \ldots b_8}] = 42\delta_{[b_1b_2b_3}^{a_1a_2a_3}\bar{Z}_{b_4\ldots b_8]} $$
$$ [R^{a_1a_2a_3}, \bar{Z}_{b_1 \ldots b_7,b}] = -{945\over4}\delta_{[b_1b_2b_3}^{a_1a_2a_3}\bar{Z}_{b_4\ldots b_7]b}-{945\over4}\delta_{[b_1b_2|b|}^{a_1a_2a_3}\bar{Z}_{b_3\ldots b_7]} \eqno(A.1.5) $$
Finally, the commutators with the negative level $E_{11}$ generators are:
$$ [R_{a_1a_2a_3}, \bar{P}^a] = -3\delta^a_{[a_1}\bar{Z}_{a_2a_3]}, \ \ \ \ [R_{a_1a_2a_3}, \bar{Z}_{a_4a_5}] = -\bar{Z}_{a_1\ldots a_5} $$
$$ [R_{a_1a_2a_3}, \bar{Z}_{b_1\ldots b_8}] = -\bar{Z}_{b_1\ldots b_5a_1a_2a_3} -\bar{Z}_{b_1\ldots b_5[a_1a_2,a_3]} $$
$$ [R_{a_1\ldots a_6}, \bar{P}^a] = -3\delta^a_{[a_1}\bar{Z}_{a_2\ldots a_6]} \ \ \ \ [R_{a_1\ldots a_6}, \bar{Z}_{b_1b_2}] = -\bar{Z}_{b_1b_2a_1\ldots a_6} -\bar{Z}_{b_1b_2[a_1\ldots a_5,a_6]} $$
$$[R_{a_1\ldots a_8,a},\bar{P}^b] = {4\over3}\delta_a^b\bar{Z}_{a_1\ldots a_8} - {4\over3}\delta^b_{[a_1}\bar{Z}_{a_2\ldots a_8]a} - {4\over3}\delta^b_{[a_1}\bar{Z}_{a_2\ldots a_8],a} \eqno(A.1.6) $$
ÔøΩ
\medskip
{\bf Appendix A.2 D=5}
\medskip
The $E_{11}$ algebra in its five dimensional decomposition will be published elsewhere. In this appendix we give the commutators of the $E_{11}$ generators with the $l_1$ generators generators of the $GL(5)\times Usp(8)$.
The commutators of $l_1$ generators with $E_{11}$ generators are
$$ [K^a{}_b, P_c] = -\delta^a_cP_b + {1\over2}\delta^a_bP_c, \ [K^a{}_b, Z^{\alpha_1\alpha_2}] = {1\over2}\delta^a_bZ^{\alpha_1\alpha_2}, \ [K^a{}_b, Z^{c\alpha_1\alpha_2}] = \delta^c_bZ^{a\alpha_1\alpha_2} + {1\over 2}\delta^a_bZ^{c\alpha_1\alpha_2} $$
$$ [R^{(\alpha_1\alpha_1)}, P_a] = 0, \ [R^{(\alpha_1\alpha_2)}, Z^{\beta_1\beta_2} = 2\Omega^{(\alpha_1[\beta_1}Z^{\alpha_2)\beta_2]} , $$
$$ [R^{(\alpha_1\alpha_2}, Z^{a\beta_1\beta_2}] = 2\Omega^{(\alpha_1[\beta_1}Z^{a\alpha_2)\beta_2]}, \ [R^{\alpha_1\ldots \alpha_4}, P_a] = 0 , $$
$$ [R^{\alpha_1\ldots \alpha_4}, Z^{\beta_1\beta_2}] = \Omega^{[\alpha_1\alpha_2}\Omega^{\alpha_3[\beta_1}Z^{\alpha_4]\beta_2]} + \Omega^{[\alpha_1[\beta_1}\Omega^{\alpha_2\beta_2]}Z^{\alpha_3\alpha_4]} $$
$$ - {1\over4}\Omega^{\beta_1\beta_2}\Omega^{[\alpha_1\alpha_2}Z^{\alpha_3\alpha_4]} - {1\over 12}\Omega^{[\alpha_1\alpha_2}\Omega^{\alpha_3\alpha_4]}Z^{\beta_1\beta_2} , $$
$$ [R^{\alpha_1\ldots \alpha_4}, Z^{a\beta_1\beta_2}] = -( \Omega^{[\alpha_1\alpha_2}\Omega^{\alpha_3[\beta_1}Z^{a\alpha_4]\beta_2]} + \Omega^{[\alpha_1[\beta_1}\Omega^{\alpha_2\beta_2]}Z^{a\alpha_3\alpha_4]} $$
$$ - {1\over4}\Omega^{\beta_1\beta_2}\Omega^{[\alpha_1\alpha_2}Z^{a\alpha_3\alpha_4]} - {1\over 12}\Omega^{[\alpha_1\alpha_2}\Omega^{\alpha_3\alpha_4]}Z^{a\beta_1\beta_2}) , \eqno(A.2.1) $$
The commutators with the positive level generators are
$$ [R^{a\alpha_1\alpha_2}, P_b] = \delta^a_bZ^{\alpha_1\alpha_2}, \ [R^{a_1a_2\alpha_1\alpha_2}, P_a] = -2\delta_b^{[a_1}Z^{a_2]\alpha_1\alpha_2} , $$
$$ [ R^{a\alpha_1\alpha_2}, Z^{\beta_1\beta_2}] = 4\Omega^{[\alpha_1[\beta_1}Z^{a\alpha_2]\beta_2]} - {1\over2}\Omega^{\beta_1\beta_2}Z^{a\alpha_1\alpha_2} - {1\over 2} \Omega^{\alpha_1\alpha_2}Z^{a\beta_1\beta_2} . \eqno(A.2.2) $$
The commutators with the negative level generators are
$$ [R_{a\alpha_1\alpha_2}, P_b] = 0, \ [R_{a\alpha_1\alpha_2}, Z^{\beta_1\beta_2}] = 2 (\delta^{\beta_1\beta_2}_{\alpha_1\alpha_2} + {1\over8}\Omega_{\alpha_1\alpha_2}\Omega^{\beta_1\beta_2})P_a, $$
$$ [R_{a\alpha_1\alpha_2}, Z^{b\beta_1\beta_2}] = 4\delta^b_a(\Omega_{[\alpha_1\gamma}\delta_{\alpha_2]}^{[\beta_1}Z^{\beta_2]\gamma} + {1\over8}\Omega_{\alpha_1\alpha_2}Z^{\beta_1\beta_2} - {1\over8}\Omega^{\beta_1\beta_2}\Omega_{\alpha_1\gamma_1}\Omega_{\alpha_2\gamma_2}Z^{\gamma_1\gamma_2}) , $$
$$ [R_{a_1a_2\alpha_1\alpha_2},Z^{b\beta_1\beta_2}] = 4(\delta^{\beta_1\beta_2}_{\alpha_1\alpha_2} + {1\over8}\Omega_{\alpha_1\alpha_2}\Omega^{\beta_1\beta_2})\delta^b_{[a_1}P_{a_2]} . \eqno(A.2.3) $$
The Cartan involution acts on the generators of $E_{11}$ and $l_1$ as
$$ I_c(K^a{}_b) = - K^b{}_a, \ \ I_c(R^{(\alpha_1\alpha_2)}) = R^{(\alpha_1\alpha_2)}, $$
$$ I_c(R^{\alpha_1\ldots \alpha_4}) = - R^{\alpha_1\ldots \alpha_4}, \ \ I_c(R^{a\alpha_1\alpha_2}) = - \Omega^{\alpha_1\beta_1}\Omega^{\alpha_2\beta_2}R_{a\beta_1\beta_2}, $$
$$ I_c(P_c) = - \bar{P}^c, \ \ I_c(Z^{\alpha_1\alpha_2}) = - \Omega^{\alpha_1\beta_1}\Omega^{\alpha_2\beta_2}\bar{Z}_{\beta_1\beta_2}, \ \ I_c(Z^{a\alpha_1\alpha_2}) = - \Omega^{\alpha_1\beta_1}\Omega^{\alpha_2\beta_2}\bar{Z}_{a\beta_1\beta_2} . \eqno(A.2.4) $$
Thus, the commutators of the level zero$E_{11}$ generators with the $\bar{l}_1$ algebra are
$$ [K^a{}_b, \bar{P}^c] = \delta^c_b\bar{P}^a - {1\over2}\delta^a_b\bar{P}^c, \ \ [K^a{}_b, \bar{Z}_{\alpha_1\alpha_2}] = -{1\over2}\delta^a_b\bar{Z}_{\alpha_1\alpha_2}, \ \ [K^a{}_b, \bar{Z}_{c\alpha_1\alpha_2}] = -\delta^c_b\bar{Z}_{a\alpha_1\alpha_2} - {1\over2}\delta^a_b\bar{Z}_{c\alpha_1\alpha_2} , $$
$$ [ R^{(\alpha_1\alpha_2)}, \bar{P}^a]=0, \ \ [ R^{(\alpha_1\alpha_2)}, \bar{Z}_{\gamma_1\gamma_2}] = 2\delta^{(\alpha_1}_{[\gamma_1}\Omega^{\alpha_2)\kappa}\bar{Z}_{\kappa\gamma_2]} $$
$$ [ R^{(\alpha_1\alpha_2)}, \bar{Z}_{a\gamma_1\gamma_2}] = 2\delta^{(\alpha_1}_{[\gamma_1}\Omega^{\alpha_2)\kappa}\bar{Z}_{a\kappa\gamma_2]} , \ \ [R^{\alpha_1\ldots \alpha_4}, \bar{P}^a] = 0 $$
$$ [ R^{\alpha_1\ldots \alpha_4}, \bar{Z}_{\kappa_1\kappa_2}] = - (\Omega^{[\alpha_1\alpha_2}\delta^{\alpha_3}_{[\kappa_1}\Omega^{\alpha_4]\gamma}\bar{Z}_{\gamma\kappa_2]} + \delta^{[\alpha_1}_{[\kappa_1}\delta^{\alpha_2}_{\kappa_2]}\Omega^{\alpha_3\gamma_1}\Omega^{\alpha_4]\gamma_2}\bar{Z}_{\gamma_1\gamma_2} $$
$$ - {1\over4} \Omega_{\kappa_1\kappa_2}\Omega^{[\alpha_1\alpha_2}\Omega^{\alpha_3\gamma_1}\Omega^{\alpha_4]\gamma_2}\bar{Z}_{\gamma_1\gamma_2} - {1\over12}\Omega^{[\alpha_1\alpha_2}\Omega^{\alpha_3\alpha_4]}\bar{Z}_{\kappa_1\kappa_2}) , $$
$$ [R^{\alpha_1\ldots \alpha_4}, \bar{Z}_{a\kappa_1\kappa_2}] = (\Omega^{[\alpha_1\alpha_2}\delta^{\alpha_3}_{[\kappa_1}\Omega^{\alpha_4]\gamma}\bar{Z}_{a\gamma\kappa_2]} + \delta^{[\alpha_1}_{[\kappa_1}\delta^{\alpha_2}_{\kappa_2]}\Omega^{\alpha_3\gamma_1}\Omega^{\alpha_4]\gamma_2}\bar{Z}_{a\gamma_1\gamma_2} $$
$$ - {1\over4} \Omega_{\kappa_1\kappa_2}\Omega^{[\alpha_1\alpha_2}\Omega^{\alpha_3\gamma_1}\Omega^{\alpha_4]\gamma_2}\bar{Z}_{a\gamma_1\gamma_2} - {1\over12}\Omega^{[\alpha_1\alpha_2}\Omega^{\alpha_3\alpha_4]}\bar{Z}_{a\kappa_1\kappa_2}) . \eqno(A.2.5) $$
The commutators of the positive level $E_{11}$ generators with the $\bar{l}_1$ representation are
$$ [ R^{a\alpha_1\alpha_2}, \bar{P}^b] = 0, \ \ [ R^{a\alpha_1\alpha_2}, \bar{Z}_{\beta_1\beta_2}] = -2(\delta^{\alpha_1\alpha_2}_{\beta_1\beta_2} + {1\over8}\Omega^{\alpha_1\alpha_2}\Omega_{\beta_1\beta_2})\bar{P}^a $$
$$ [R^{a\alpha_1\alpha_2}, \bar{Z}_{b\beta_1\beta_2}] = 4\delta^a_b(\Omega^{[\alpha_1\gamma}\delta^{\alpha_2]}_{[\beta_2}\bar{Z}_{\beta_2]\gamma} + {1\over8}\Omega^{\alpha_1\alpha_2}\bar{Z}_{\beta_1\beta_2} - {1\over8}\Omega_{\beta_1\beta_2}\Omega^{\alpha_1\gamma_1}\Omega^{\alpha_2\gamma_2}\bar{Z}_{\gamma_1\gamma_2}) , $$
$$ [R^{a_1a_2\alpha_1\alpha_2}, \bar{Z}_{b\beta_1\beta_2}] = 4(\delta^{\alpha_1\alpha_2}_{\beta_1\beta_2} + {1\over8}\Omega^{\alpha_1\alpha_2}\Omega_{\beta_1\beta_2})\delta_b^{[a_1}\bar{P}^{a_2]} . \eqno(A.2.6) $$
Finally, the commutators of the $\bar{l}_1$ algebra with the negative level $E_{11}$ generators are
$$ [R_{a\alpha_1\alpha_2}, \bar{P}^b] = -\delta^b_a\bar{Z}_{\alpha_1\alpha_2}, \ \ [R_{a_1a_2\alpha_1\alpha_2}, \bar{P}^b] = 2\delta^b_{[a_1}\bar{Z}_{a_1]\alpha_1\alpha_2} , $$
$$ [R_{a\alpha_1\alpha_2}, \bar{Z}_{\beta_1\beta_2}] = - (4\Omega_{\alpha_1[\beta_1}\bar{Z}_{a\alpha_2]\beta_2]} - {1\over2}\Omega_{\beta_1\beta_2}\bar{Z}_{a\alpha_1\alpha_2} - {1\over2}\Omega_{\alpha_1\alpha_2}\bar{Z}_{a\beta_1\beta_2} ) . \eqno(A.2.7) $$
\medskip
{\bf Appendix A.3 D=4}
\medskip
The commutators of the $E_{11}\otimes_sl_1$  decomposed algebra into representations of $GL(4)\times SL(8)$ can be found in [16]. The commutators of the level zero $E_{11}$ generators with the $l_1$ generators are
$$ [K^a{}_b, P_c] = -\delta^a_cP_b + {1\over 2}\delta^a_bP_c, \ [K^a{}_b, Z^{I_1I_2}] = {1\over2}\delta^a_bZ^{I_1I_2}, $$
$$ [K^a{}_b, Z_{I_1I_2}] = {1\over2}\delta^a_bZ_{I_1I_2}, \ [K^a{}_b, Z^c] = \delta^c_bZ^a + {1\over2}\delta^a_bZ^c, $$
$$ [R^I{}_J, P_c] = 0, \ [R^I{}_J, Z^{I_1I_2}] = 2\delta_J^{[I_1}Z^{|I|I_2]} - {1\over4}\delta^I_JZ^{I_1I_2}, $$
$$ [ R^I{}_J, P_c] = 0, \ [R^I{}_J, Z_{I_1I_2}] = -2\delta^I_{[I_1}Z_{|J|I_2]} + {1\over4}\delta^I_JZ_{I_1I_2}, \ [R^I{}_J, Z^a] = 0, $$
$$ [R^{I_1\ldots I_4}, P_a] = 0, \ [R^{I_1\ldots I_4}, Z^{J_1J_2}] = {1\over24}\varepsilon^{I_1\ldots I_4J_1\ldots J_4}Z_{J_3J_4}, $$
$$ [ R^{I_1\ldots I_4},Z_{J_1J_2}] = \delta_{J_1J2_2}^{[I_1I_2}Z^{I_3I_4]}, \ [R^{I_1\ldots I_4}, Z^a] = 0 . \eqno(A.3.1) $$
The commutators of the positve level $E_{11}$ generators with the $l_1$ generators are
$$ [R^{aI_1I_2}, P_b] = \delta^a_bZ^{I_1I_2}, R^a{}_{I_1I_2}, P_b] = \delta^a_bZ_{I_1I_2}, $$
$$ [R^a{}_{I_1I_2}, Z^{J_1J_2}] = \delta^{J_1J_2}_{I_1I_2}Z^a , \ [R^{aI_1I_2}, Z_{J_1J_2}] = -\delta_{J_1J_2}^{I_1I_2}Z^a . \eqno(A.3.2) $$
The commutators with the negative level generators are
$$ [\tilde{R}_{aI_1I_2}, P_b] = 0, \ [\tilde{R}_{aI_1I_2}, Z^{J_1J_2}] = 2\delta^{I_1I_2}_{J_1J_2}P_a, [\tilde{R}_{aI_1I_2}, Z_{J_1J_2}] = 0, $$
$$ [\tilde{R}_a{}^{I_1I_2}, P_b] = 0, \ [\tilde{R}_a{}^{I_1I_2},Z^{J_1J_2}] = 0, \ [\tilde{R}_a{}^{I_1I_2},Z_{J_1J_2}] = -2\delta_{I_1I_2}^{J_1J_2}P_a , $$
$$ [\tilde{R}_{aI_1I_2}, Z^b] = -2\delta^b_aZ_{I_1I_2}, \ [ \tilde{R}_a{}^{I_1I_2}, Z^b] = -2\delta_a^bZ^{I_1I_2}. \eqno(A.3.3) $$
The Cartan involution acts on the $E_{11}$ generators as
$$I_c(K^a{}_b) = -K^b{}_a, \ \ I_c(R^I{}_J) = -R^J{}_I, \ \ I_c(R^{I_1\ldots I_4}) = -\star R^{I_1\ldots I_4} \equiv -{1\over 4!}\epsilon^{I_1\ldots I_4J_1\ldots J_4}R^{J_1\ldots J_4}, $$
$$ I_c(R^{aI_1I_2}) = -\tilde{R}_{aI_1I_2}, \ \ I_c(R^a{}_{I_1I_2}) = \tilde{R}_a{}^{I_1I_2} , \eqno(A.3.4) $$
and on the $l_1$ representation as
$$ I_c(P_c) = -\bar{P}^c, \ \ I_c(Z^{I_1I_2}) = -\bar{Z}_{J_1J_2}, $$
$$ I_c(Z_{I_1I_2}) = - \bar{Z}^{J_1J_2} , \ \ I_c(Z^c) = - \bar{Z}_c . \eqno(A.3.5) $$
The commutators of the level zero $E_{11}$ algebra with the $\bar{l}_1$ representation are
$$ [K^b{}_a, \bar{P}^c] = -\delta^c_a\bar{P}^b + {1\over 2}\delta^b_a\bar{P}^c, \ [K^b{}_a, \bar{Z}_{I_1I_2}] = - {1\over2}\delta_a^b\bar{Z}_{I_1I_2}, $$
$$ [K^b{}_a, \bar{Z}^{I_1I_2}] = -{1\over2}\delta_a^b\bar{Z}^{I_1I_2}, \ [K^b{}_a, \bar{Z}_c] = -\delta_c^b\bar{Z}_a - {1\over2}\delta_a^b\bar{Z}_c, $$
$$ [R^I{}_J, \bar{P}^c] = 0, \ [R^I{}_J, \bar{Z}_{I_1I_2}] = -2\delta^J_{[I_1}\bar{Z}_{|I|I_2]} + {1\over4}\delta_I^J\bar{Z}_{I_1I_2}, $$
$$ [R^I{}_J, \bar{Z}^{I_1I_2}] = 2\delta_I^{[I_1}\bar{Z}^{|J|I_2]} - {1\over4}\delta_I^J\bar{Z}^{I_1I_2}, \ [R^I{}_J, \bar{Z}_a] = 0, $$
$$ [R^{I_1\ldots I_4}, \bar{P}^a] = 0, \ [R^{I_1\ldots I_4}, \bar{Z}_{J_1J_2}] = -\delta_{J_1J2_2}^{[I_1I_2}Z^{I_3I_4]} , $$
$$ [ R^{I_1\ldots I_4},\bar{Z}^{J_1J_2}] = -{1\over24}\varepsilon^{I_1\ldots I_4J_1\ldots J_4}Z_{J_3J_4}, \ [R^{I_1\ldots I_4}, \bar{Z}_a] = 0. \eqno(A.3.6) $$
The commutators of the positve level $E_{11}$ generators with the $l_1$ generators are
$$ [R^{aI_1I_2}, \bar{P}^b] = 0, \ [R^{aI_1I_2}, \bar{Z}_{J_1J_2}] = -2\delta^{I_1I_2}_{J_1J_2}\bar{P}_a, \ [R^{aI_1I_2}, \bar{Z}^{J_1J_2}] = 0, $$
$$ [R^a{}_{I_1I_2}, \bar{P}^b] = 0, \ [R^a{}_{I_1I_2},\bar{Z}_{J_1J_2}] = 0, \ [R^a{}_{I_1I_2},\bar{Z}^{J_1J_2}] = -2\delta_{I_1I_2}^{J_1J_2}\bar{P}^a , $$
$$ [R^{aI_1I_2}, \bar{Z}_b] = 2\delta^a_b\bar{Z}^{I_1I_2}, \ [ R^a{}_{I_1I_2}, \bar{Z}_b] = -2\delta_b^a\bar{Z}_{I_1I_2}. \eqno(A.3.7) $$
The commutators with the negative level generators are
$$ [\tilde{R}_{aI_1I_2}, \bar{P}^b] = -\delta^b_a\bar{Z}_{I_1I_2}, \ [\tilde{R}_a{}^{I_1I_2}, \bar{P}^b] = \delta^b_a\bar{Z}^{I_1I_2}, $$
$$ [\tilde{R}_a{}^{I_1I_2}, \bar{Z}_{J_1J_2}] = -\delta_{J_1J_2}^{I_1I_2}\bar{Z}_a , \ [\tilde{R}_{aI_1I_2}, \bar{Z}^{J_1J_2}] = -\delta^{J_1J_2}_{I_1I_2}\bar{Z}_a . \eqno(A.3.8) $$
\medskip
{\bf Appendix A.4 $A_1^{+++}$}
\medskip
The commutators of the $A_1^{+++}$ can be found in reference [18]. The commutators of the level zero generators of $GL(4)$ with the $l_1$ algebra are
$$ [K^a{}_b, P_c] = -\delta^a_cP_b + {1\over2}\delta^a_bP_c, \ [K^a{}_b,Z^c] = \delta^c_bZ^a + {1\over2}\delta^a_bZ^c, $$
$$ [ K^a{}_b, Z^{cde}] = \delta_b^cZ^{ade} + \delta_b^dZ^{cae} + \delta^e_bZ^{cda} + {1\over2}\delta^a_bZ^{cde} , $$
$$ [K^a{}_b, Z^{cd,e}] = \delta^c_bZ^{ad,e} + \delta^d_bZ^{ca,e} + \delta^e_bZ^{cd,a} + {1\over2}\delta^a_bZ^{cd,e} , \eqno(A.4.1) $$
The commutators with the positive level generators are
$$ [R^{ab}, P_c] = \delta_c^{(a}Z^{b)}, \ [R^{ab}, Z^c] = Z^{abc} + Z^{c(a,b)} \eqno(A.4.2) $$
The commutators with the negative level generators are
$$[R_{ab}, P_c] = 0, \ [R_{ab}, Z^c] = 2\delta^c_{(a}P_{b)} , $$
$$ [R_{ab}, Z^{cde}] = {2\over3}(\delta^{cd}_{(ab)}Z^e + \delta^{de}_{(ab)}Z^c + \delta^{ec}_{(ab)}Z^d) , $$
$$ [ R_{ab}, Z^{cd,e}] = {4\over3}(\delta^de_{(ab)}Z^c - \delta^{ce}_{(ab)}Z^d) . \eqno(A.4.3) $$
The Cartan involution acts on the $E_{11}$ generators as
$$ I_c(K^a{}_b) = -K^b{}_a, \ \ I_c(R_{ab}) = -R^{ab}, \ \ I_c(R^{ab,cd}) = R_{ab,cd} , $$
and on the $l_1$ representation as
$$ I_c(P_c) = -\bar{P}^c , \ \ I_c(Z^c) = -\bar{Z}_c , $$
$$ I_c(Z^{cde}) = - \bar{Z}_{cde}, \ \ I_c(Z^{cd,e}) = - \bar{Z}_{cd,e} . \eqno(A.4.4) $$
The commutators of the level zero generators with the $\bar{l}_1$ algebra are
$$ [K^b{}_a, \bar{P}^c] = \delta^c_a\bar{P}^b - {1\over2}\delta^b_a\bar{P}^c, \ [K^b{}_a,\bar{Z}^c] = -\delta^b_c\bar{Z}^a -{1\over2}\delta^b_a\bar{Z}_c , $$
$$ [ K^b{}_a, \bar{Z}_{cde}] = -\delta^b_c\bar{Z}_{ade} - \delta^b_d\bar{Z}_{cae} - \delta^b_e\bar{Z}_{cda} - {1\over2}\delta_a^b\bar{Z}_{cde} , $$
$$ [K^b{}_a, \bar{Z}_{cd,e}] = - \delta^b_c\bar{Z}_{ad,e} - \delta^b_d\bar{Z}_{ca,e} - \delta^b_eZ_{cd,a} - {1\over2}delta^b_a\bar{Z}_{cd,e}, $$
$$ [ K^b{}_a, \bar{Z}_{cd,e}] = \delta^b_c\bar{Z}_{ad,e} - \delta^b_d\bar{Z}_{ca,e} - \delta^b_e\bar{Z}_{cd,a} - {1\over2} \delta^b_a\bar{Z}_{cd,e} . \eqno(A.4.5) $$
Then the commutators with the positive level generators of $GL(4)$ and $\bar{l}_1$ are
$$ [ R^{ab}, \bar{P}^c] = 0 , \ \ [R^{ab}, \bar{Z}_c] = -2\delta_c^{(a}\bar{P}^{b)} , $$
$$ [ R^{ab}, \bar{Z}_{cde}] = - {2\over 3}(\delta_{cd}^{(ab)}\bar{Z}_e + \delta_{de}^{(ab)}\bar{Z}_c + \delta_{ec}^{(ac)}\bar{Z}_d) , $$
$$ [ R^{ab}, \bar{Z}_{cd,e} ] = -{4\over3}(\delta_{de}^{(ab)}\bar{Z}_c - \delta_{ce}^{(ab)}\bar{Z}_d) . \eqno(A.4.6) $$
Finally, the commutators with the negative level generators of $GL(4)$ and $\bar{l}_1$ are
$$ [R_{ab}, \bar{P}^c] = -\delta^c_{(a}\bar{Z}_{b)}, \ \ [R_{ab}, \bar{Z}^c] = -\bar{Z}_{abc} - \bar{Z}_{c(a,b)} , $$
$$ [ R_{ab,cd},\bar{P}^e] = -\delta^e_{[a}\bar{Z}_{b]cd} + {1\over 4}(\delta^e_a\bar{Z}_{b(c,d)} - \delta^e_b\bar{Z}_{a(c,d)} - {3\over8}(\delta^e_c\bar{Z}_{ab,d} + \delta^e_a\bar{Z}_{ab,c}) . \eqno(A.4.7) $$


\medskip
{\bf {References}}
\medskip
\item{[1]} E. Cremmer and B. Julia,
{\it The $N=8$ supergravity theory. I. The Lagrangian},
Phys.\ Lett.\ {\bf 80B} (1978) 48.
\item {[2]} B.\ Julia, {\it  Group Disintegrations},
in {\it Superspace
Supergravity}, p.\ 331,  eds.\ S.W.\ Hawking  and M.\ Ro\v{c}ek,
Cambridge University Press (1981);
 E. Cremmer, {\it Dimensional Reduction In
Field Theory And Hidden Symmetries In Extended Supergravity},  Published
in Trieste Supergravity School 1981, 313;  {\it Supergravities In 5
Dimensions}, in {\it Superspace
Supergravity}, p.\ 331,  eds.\ S.W.\ Hawking  and M.\ Ro\v{c}ek,
Cambridge University Press (1981).
\item{[{3}]} J, Schwarz and P. West,
{\it  Symmetries and Transformation of Chiral
$N=2$ $D=10$ Supergravity},
\item{[4]} P. West, {\it $E_{11}$ and M Theory}, Class. Quant. 
Grav.  {\bf 18}
(2001) 4443, {\tt arXiv:hep-th/ 0104081};
\item{[5]} P. West, {\it $E_{11}$, SL(32) and Central Charges},
Phys. Lett. {\bf B 575} (2003) 333-342, {\tt hep-th/0307098}
\item{[6]} P. West, {\it A brief review of E theory}, Proceedings of Abdus Salam's 90th  Birthday meeting, 25-28 January 2016, NTU, Singapore, Editors L. Brink, M. Duff and K. Phua, World Scientific Publishing and IJMPA, {\bf Vol 31}, No 26 (2016) 1630043,  arXiv:1609.06863,
\item{[7]} A. Tumanov and  P. West, {\it E11 and exceptional field theory}, Int.J.Mod.Phys.A31, (2016) no. 12, 1650066, arXiv:1507.08912.
\item{[8]} A. Borisov and V. Ogievetsky,  {\it Theory of dynamical affine and conformal  symmetries as the theory of the gravitational field},
Teor. Mat. Fiz. 21 (1974) 32
\item{[9]} P. West, {\it Hidden superconformal symmetries of 
M-theory}, {\bf JHEP 0008} (2000) 007, {\tt arXiv:hep-th/0005270}.
\item{[10]} P. West, Introduction to Strings and Branes, Cambridge
University Press, June 2012.
\item{[11]}  P. West, {\it Generalised Geometry, eleven dimensions
and $E_{11}$}, JHEP 1202 (2012) 018, arXiv:1111.1642. 
\item{[12]} A. Tumanov and P. West, {\it E11 must be a symmetry of strings and branes },  Phys. Lett. {\bf  B759 } (2016),  663, arXiv:1512.01644.
\item{[13]} A. Tumanov and P. West, {\it E11 in 11D}, Phys.Lett. B758 (2016) 278, arXiv:1601.03974.
\item{[14]} D. Berman, H. Godazgar, M. Perry and P. West, {\it Duality Invariant Actions and Generalised Geometry}, JHEP 1202 (2012) 108, arXiv:1111.0459
\item{[15]} F. Riccioni and P. West, {\it E(11)-extended spacetime
and gauged supergravities}, JHEP {\bf 0802} (2008) 039, arXiv:0712.1795.
\item{[16]} P. West, {\it  E11, Generalised space-time and equations of motion in four dimensions}, JHEP 1212 (2012) 068, arXiv:1206.7045.
\item{[17]}  N. Lambert and P. West, {\it Coset Symmetries in Dimensionally Reduced Bosonic String Theory}, Nucl.Phys. B615 (2001) 117-132, hep-th/0107209.
\item{[18]} A. Tumanov and P. West, {\it Generalised vielbeins and non-linear realisations }, JHEP 1410 (2014) 009,  arXiv:1405.7894.
\item{[19]} P. West, {\it Generalised Space-time and Gauge Transformations}, JHEP 1408 (2014) 050, arXiv:1403.6395.
\item{[20]} V. Kac, {\it Infinite Dimensional Lie algebras}, Birkhauser, 1983. See in particular proposition 9.4 page 106.
\item{[21]} F. Englert and L. Houart, {\it G+++ Invariant Formulation of Gravity and M-Theories: Exact BPS Solutions }, JHEP0401:002,2004,  arXiv:hep-th/0311255.


\end

\item{[27]} A. Coimbra, C. Strickland-Constable and  D.  Waldram,
{\it Supergravity as Generalised Geometry I: Type II Theories},
arXiv:1107.1733; {\it $E_{d(d)} \times {R}^+$ Generalised Geometry,
Connections and M theory}, arXiv:1112.3989. A. Coimbra, C. Strickland-Constable and  D.  Waldram, {\it  Supergravity as Generalised Geometry II: $E_{d(d)} \times {R}^+$ and M theory} , arXiv:1212.1586; P. Pacheco and  D.  Waldram, {\it M-theory, exceptional generalised geometry and superpotentials},  JHEP0809 (2008) 123,  arXiv:0804.1362
\item{[28]} N. Hitchin, Generalized Calabi-Yau manifolds,
 Q. J. Math.  {\bf 54}  (2003), no. 3, 281,
math.DG/0209099;  {\it Brackets, form and
invariant functionals}, math.DG/0508618;  M. Gualtieri, {\it Generalized complex geometry}, PhD Thesis
(2004), math.DG/0401221v1.